\documentclass[fleqn]{article}
\usepackage[left=2.5cm,top=2.5cm,right=2.5cm,bottom=2.5cm]{geometry}
\usepackage{amsmath}
\usepackage{amsfonts}
\usepackage{amssymb}
\usepackage{graphicx}
\usepackage{caption}
\usepackage{subfigure}
\usepackage{cleveref}
\usepackage{sectsty}
\usepackage{xcolor}
\usepackage{multirow}
\usepackage{float}
\usepackage{braket} 
\usepackage{bigints}
\usepackage{enumitem}

\def\tu{\tilde{u}}
\def\baru{\bar{V}_x}

\def\barrho{\bar{\rho}}

\def\Re{\mbox{Re}}
\def\Ma{\mbox{Ma}}
\def\Pr{\mbox{Pr}}

\def\kx{k_x}

\def\trho{\hat{\rho}}

\def\tT{\hat{T}}
\def\tp{\hat{p}}

\def\tT{\tilde{T}}

\def\barT{\bar{T}}

\def\kxsq{k_x^2}

\def\barrho{\bar{\rho}}

\def\barT{\bar{T}}
\def\barp{\bar{p}}
\def\barkappa{\bar{\kappa}}
\def\trho{\hat{\rho}}
\def\tu{\hat{v}_x}

\def\tv{\hat{v}_y}

\def\tuzero{\hat{u}_0}

\def\tvzero{\hat{v}_0}

\def\cone{c_1}

\def\czero{c_0}
\def\tp{\hat{p}}
\def\tpzero{\hat{p}_0}

\def\tT{\hat{T}}
\def\Re{\mbox{Re}}
\def\Ma{\mbox{Ma}}
\def\Pr{\mbox{Pr}}

\def\cR{c_R}
\def\cI{c_I}

\def\barx{\bar{x}}
\def\bary{\bar{y}}

\def\barus{\bar{U}_{x}}
\def\barvs{\bar{U}_{y}}

\def\pus{u_x^\prime}
\def\pvs{u_y^\prime}
\def\barmu{\bar{\mu}}

\def\tus{\hat{u}_x}

\def\tvs{\hat{u}_y}

\def\tuszero{\hat{u}_x^{(0)}}
\def\tvszero{\hat{u}_y^{(0)}}
\def\tpszero{\hat{p}_s^{(0)}}
\def\tTszero{\hat{T}_s^{(0)}}
\def\tuzero{\hat{v}_x^{(0)}}
\def\tvzero{\hat{v}_y^{(0)}}
\def\tpzero{\hat{p}^{(0)}}
\def\czero{c^{(0)}}
\def\tusone{\hat{u}_x^{(1)}}
\def\tvsone{\hat{u}_y^{(1)}}
\def\tpsone{\hat{p}_s^{(1)}}

\def\cone{c^{(1)}}

\def\tvsa{\hat{u}_y^\ast}
\def\tvsazero{\hat{u}_y^{(0) \ast}}
\def\tvsaone{\hat{u}_y^{(1) \ast}}
\def\czerosq{c^{(0) 2}}
\def\cRzero{c_R^{(0)}}
\def\cIzero{c_I^{(0)}}
\def\cIone{c_I^{(1)}}

\def\tsigmaxxzero{\hat{\sigma}_{xx}^{(0)}}
\def\tsigmaxyzero{\hat{\sigma}_{xy}^{(0)}}
\def\tsigmayyzero{\hat{\sigma}_{yy}^{(0)}}

\def\tTzero{\hat{T}^{(0)}}
\def\barsigmatw{\bar{\sigma}_{tw}}

\def\tuazero{\hat{v}_x^{(0) \ast}}
\def\tvazero{\hat{v}_y^{(0) \ast}}
\def\divzero{\mathrm{div}^{(0)}}
\def\divazero{\mathrm{div}^{(0) \ast}}
\def\ttauxyzero{\hat{\tau}_{xy}^{(0)}}
\def\ttauxyazero{\hat{\tau}_{xy}^{(0) \ast}}
\def\tPhizero{\hat{\Phi}^{(0)}}

\begin{document}

\title{\textbf{Stability of gas flow past viscoelastic compliant solid} 
}%

\author{Mandeep Deka\thanks{email: mandeep\_deka@iitg.ac.in} \\\small{Mechanical Engineering Department},\\ \small{Indian Institute of Technology Guwahati, India.} }


\date{}

\maketitle

\begin{abstract}
Stability of a high-speed gas flow past a compliant solid is impacted by two distinct features: high solid-to-fluid density ratio ($\rho_r$), and flow compressibility when flow speeds are comparable to acoustic speed. This study investigates the linear stability of a shear-driven compressible gas flow past a compliant substrate modelled as a continuum Neo-Hookean solid. Numerical solutions of the eigenvalue problem reveal that at high density ratios, the dominant instabilities are the elastic shear-waves of the solid. Our study shows that flow compressibility exerts a non-monotonic effect on the growth rate of the elastic modes; the growth rate increases with increase in Mach number up to $\Ma \approx 2$ before subsequently decreasing. Furthermore, for compressible flows, strong thermal-coupling renders the base state highly sensitive to both the solid-to-fluid thermal conductivity ratio and the substrate's bottom-surface temperature. Numerical results demonstrate that increasing the conductivity ratio destabilizes the system, whereas increasing the bottom wall temperature is stabilizing. The stability equations, analyzed in the asymptotic limit of $\rho_r\Re \gg 1$, reveals that the fluid-solid system de-couples at the leading order with the elastic modes emerging as a solution of the linear elasticity equations under free-shear condition at the interface. We derive a closed-form expression for the leading-order growth rate of the instability, which shows an excellent agreement with the numerical solution. This expression explicitly quantifies the influence of fluid stresses at the interface, which can in-turn be expressed as integrals of the flow solution isolating the distinct physical mechanisms driving the instability.
\end{abstract}

\section{Introduction}
\label{sec:intro}

The interaction of fluid flows with compliant solids span a diverse range of scientific and engineering disciplines. In biomechanics, for instance, pulmonary respiration can be modeled as gas transport through conduits lined with a viscoelastic mucosal layer \cite{grotberg2001respiratory,ref-evrensel-93,patne2021purely}. Conversely, in external aerodynamics, the perennial objective of drag reduction via viscoelastic surface coatings has remained a focal point of research ever since the early experiments by Kramer \cite{ref-kramer-60,ref-kramer-61,ref-kramer-62}. Driven by two divergent objectives - namely, delaying laminar-to-turbulent transition through instability suppression versus accelerating fluid mixing via instability onset - hydrodynamic stability analysis has provided important insights in furthering the understanding of fluid-structure interaction in these systems.

Over the past several decades, numerous studies have explored the characteristics of flows past compliant surfaces through theoretical \cite{ref-benjamin-63,ref-yeo-87,ref-kumaran-96,ref-kumaran-2021}, computational \cite{ref-lucey-92,davies1997numerical,rosti2017numerical,xia2017direct} and experimental \cite{ref-gad-84,kumaran2000spontaneous,eggert2004observations,ref-verma-2012,ref-verma-2013} approaches. 
The seminal stability studies by \cite{ref-benjamin-60,ref-benjamin-63,ref-landahl-62,ref-carpenter-85,ref-carpenter-86}, that modelled compliant surfaces as spring-backed plates, demonstrated the stabilizing property of wall compliance for the existing rigid wall modes, but also showed the destabilization potential through creation of flow-induced-structural (FISI) instabilities. While simple impedance-type models remain widely utilized due to their lower computational cost, they suffer from parameter ambiguity and fail to capture critical instabilities driven by tangential wall motion \cite{ref-shankar-2002}. This necessitates the adoption of a continuum-based framework for the compliant medium. The stability of flows past compliant surfaces modelled as viscoelastic continuum solids have been studied for shear-driven flows (\cite{kumaran1995effect,ref-shankar-2002}) and pressure-driven flows in channels and tubes (\cite{ref-shankar-2000,ref-gaurav-2009,ref-gaurav-2010}), as well as for boundary layer flows (\cite{evrensel1988response}). Although a continuum formulation significantly increases the mathematical complexity -- requiring the simultaneous solution of coupled governing equations for both the fluid and solid domains -- it ensures all physical modes that can destabilise the system are accurately captured. Crucially, the instabilities predicted by classical linear stability analyses have been validated experimentally, notably at low Reynolds numbers in Couette configurations \cite{muralikrishnan2002experimental} and at moderate Reynolds numbers in pressure-driven channel and pipe flows \cite{ref-verma-2013,ref-verma-2012}.

The simplest flow configuration of a shear-driven (Couette) flow past a viscoelastic compliant solid yields novel instabilities that do not exist in the rigid-wall counterparts. At low Reynolds numbers, the system destabilizes via a ``viscous mode'' once the wall elasticity exceeds a critical threshold \cite{kumaran1994flow}. This instability is generated due the transfer of energy from base flow to perturbations due to shear deformation at the interface. Conversely, in the high-Reynolds-number regime, two distinct classes of instabilities emerge: (a) the ``inviscid modes'' generated due to the elastic stresses in the solid (\cite{kumaran1995effect}), and, (b) a ``wall-mode'' which arises from localized shear deformation within a critical layer of thickness $O(\Re^{-1/3})$ at the fluid-solid interface (\cite{ref-srivatsan-97}), $\Re$ being the flow Reynolds number defined in section \ref{subsec:gov_eqs} (Eq. \ref{eq:nd_defs}). The wall-mode is the finite-$\Re$ extension of the viscous instability at low Reynolds number and is in general dependent only on the value of the shear stress at the wall. In contrast, the inviscid modes manifest as travelling shear waves within the elastic solid that are destabilized by the overlying fluid flow. A comprehensive review of the instabilities in internal flows bounded by compliant boundaries is provided by \cite{ref-kumaran-2021,shankar2015stability}. 
A key point to note is that the predominant study of the linear modes are carried out in the context of liquid-solid systems, whereas the corresponding studies for gas flows are relatively rare.

A defining characteristic of gas flows past compliant solids, distinguishing them from liquid-solid configurations, is the profound density mismatch between the fluid medium and the underlying solid substrate. A recent study by \cite{deka2025effect} demonstrated that an increase in solid-to-fluid density ratio ($\rho_r$) causes significant destabilization, particularly at finite Reynolds numbers ($\Re$)  relevant to biological and aerospace applications. At high $\rho_r$, the dominant mode of instability shifts from the wall mode to the elastic modes of the solid (\cite{kumaran1995effect}) with the critical Reynolds number, $\Re_c \sim \rho_r^{-1/2}$. In the biomedical context of airflow through the human respiratory tract and oral cavities, the elastic instabilities have been studied in detail by \cite{patne2021purely}, who modeled the airflow over a mucous layer -- represented as a shear-thinning viscoelastic liquid -- overlying a deformable solid substrate. Their analysis revealed that the predominant instability mechanism stems from a resonance between the ``solid elastic'' and ``liquid elastic'' modes of the system.

Besides high density ratio, another important aspect, particularly in the context of high speed external flows, is fluid compressibility. When the flow speeds are comparable to the speed of sound, the hydrodynamics and thermodynamics couple leading to new instabilities absent in incompressible flows. With an increase in flow compressibility, the nature of traveling perturbations shifts from purely hydrodynamic shear waves to high-frequency acoustic modes, referred to as ``Mack modes'' \cite{ref-mack-63,ref-mack-65,ref-mack-84,ref-mack-87}. Furthermore, high-speed compressible flows induce severe aerodynamic heating that modifies the local temperature field, thereby altering the fluid viscosity and density profiles and shifting the corresponding stability boundaries. In supersonic boundary layer, transition is primarily dictated by the second mode instability, which is known to be destabilized by wall cooling \cite{ref-mack-84}. For bounded configurations, such as compressible Couette flows, the compressible modes are typically categorised into an upstream- and a downstream-running family of acoustic waves sustained via reflections between the wall and the sonic line \cite{ref-duck-94,ref-hu-98}.  


Despite its physical relevance, investigations into compressible flows past compliant boundaries remain comparatively sparse. \cite{ref-walker-2006} utilized asymptotic triple-deck theory to analyze the modification of Tollmien-Schlichting modes in a supersonic flow over an elastic membrane. More recently, \cite{ref-dettenrieder-2022} studied the stability of a compressible boundary layer over a compliant wall governed by the Kirchhoff-Love plate model, delineating two distinct regions of the stability plane: one corresponding to a piston-like panel motion, and the other corresponding to flexural plate movement independent of Mach number. Extending classical stability theorems to compressible flows bounded by spring-backed plates, \cite{deka2025inviscid} mapped out stability bounds within the inviscid limit. The recent work of \cite{deka2025linear} investigate the stability of a compressible channel flow with compliant walls modelled as spring-backed plates, demonstrating a reciprocal stabilization effect: flow compressibility stabilizes the flow-induced structural modes, while wall elasticity conversely stabilizes the compressible modes of the fluid. 

Use of compliant coatings for compressible flows have garnered a renewed interest over the past decade due to its potential in flow control. Experiments conducted by \cite{ref-pham-2018} on shock-wave boundary layer interaction over a rubber coating showed a 50-60 \% reduction in the energy content of the separation shock oscillations, highlighting their prospect in shock-load reduction. Similarly, recent investigations by \cite{walz2023dual,ref-chakravarty-2024} also show a marked reduction in the skin friction coefficient with viscoelastic rubber implants, indicating potential for compliant coatings in drag reduction applications. 

The motivation of the present study lies in the intersection of the following research gaps -
\begin{itemize}
\item The stability of flows past compliant surfaces characterized by large solid-to-fluid density ratios remains insufficiently explored. Specifically, the characteristics of the dominant instabilities and their destabilization mechanisms are not quantitatively identified. This study addresses this gap by isolating and quantifying the unstable modes in the linear regime.
\item Prior investigations into compressible flows over compliant boundaries have predominantly relied on low-order wall response models (e.g., spring-backed plates). These simplified representations fail to capture the multi-dimensional deformation and viscous dissipation inherent to viscoelastic continuum substrates. We demonstrate that the interplay between flow compressibility and continuum wall dynamics yields stability characteristics that deviate qualitatively from the predictions of spring-backed plate type models.
\item Fluid compressibility introduces thermal coupling at the fluid-solid interface. Simple mechanical wall models are fundamentally incapable of accounting for these thermodynamic interactions, leading to potentially inaccurate stability bounds. This work explicitly incorporates a coupled energy formulation to reveal how thermal transport properties -- such as the solid-to-fluid thermal conductivity ratio and the bottom wall temperature boundary conditions -- significantly modify the stability boundaries.
\item Finally, analytical results for coupled fluid-solid systems, where both media are governed by continuum equations, are exceedingly rare. To circumvent the high computational cost of fully numerical stability analyses, we derive closed-form, asymptotic solutions for the leading-order instability in the high-density-ratio limit. These analytical expressions generalize and considerably simplify stability calculations for both incompressible and compressible gas flows past viscoelastic solids.
\end{itemize}

In this work, we study the linear stability of a compressible shear-driven gas flow past a viscoelastic solid layer using numerical eigenvalue calculations and asymptotic analyses. The remainder of the paper is organized as follows. Section \ref{sec:formulation} establishes the mathematical formulation of the coupled problem, outlining the governing continuum equations for both the fluid and solid domains, the derivation of the compressible base state, and the linearized disturbance equations. The numerical solution of the stability equations are presented in section \ref{sec:results}. The dominant instability is identified and the variation of the instability growth with relevant parameters is studied in this section. Following that in section \ref{sec:asymptotic}, an asymptotic analysis is presented in the high density ratio limit. The leading order growth rate for the dominant instability is obtained and an interpretation of the interfacial flow contributions to the instability is discussed. Finally, section \ref{sec:conclusions} summarizes the primary findings and key takeaways of this study.

\section{Formulation}
\label{sec:formulation}

\begin{figure}
\begin{center}
\begin{minipage}{.55\textwidth}
  \includegraphics[width=3.5in]{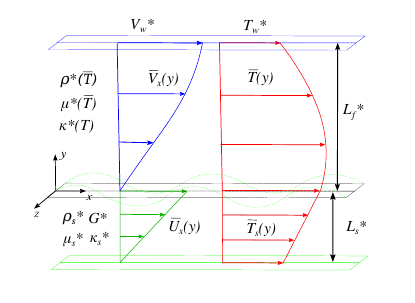}
  \caption{Schematic of a Couette flow past a viscoelastic compliant solid}
  \label{fig-schm}
\end{minipage}
\qquad
\begin{minipage}{.35\textwidth}
  \includegraphics[width=2.5in]{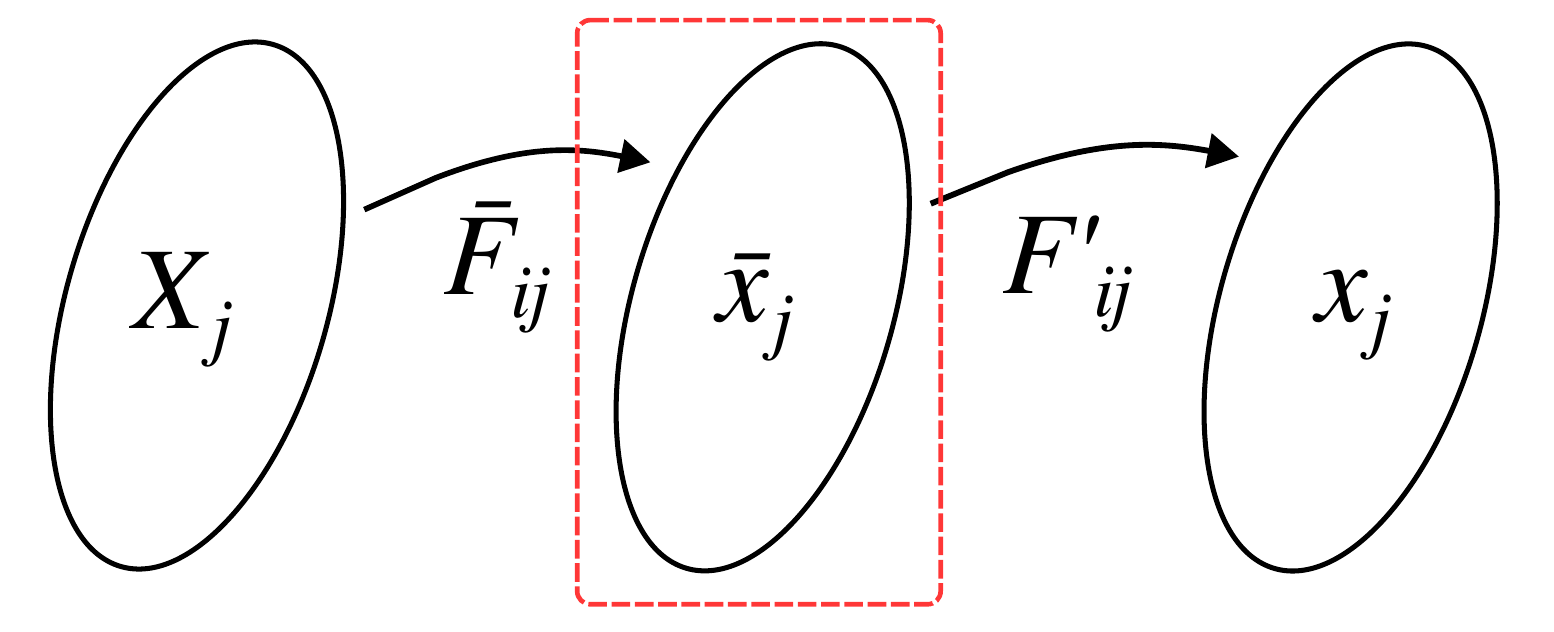}
  \caption{Schematic of the three-state configuration for the solid deformation.} 
  \label{fig-config}
\end{minipage}
\end{center}
\end{figure}

To study the stability of a gas flow past a compliant surface, we consider the canonical problem of a Couette flow with the bottom surface replaced by layer of a viscoelastic compliant solid. Figure \ref{fig-schm} shows a schematic of the configuration which comprises of an infinitely large plate and an infinitely large solid layer arranged parallelly with a fluid entrapped in between. Flow is driven by the top surface moving with a velocity $V_w^\ast$. The bottom surface is a viscoelastic compliant solid layer whose top surface interacts with the flow and the bottom surface is pinned to a rigid base. 

The thickness of the fluid and solid layers are denoted by $L_f^\ast$ and $L_s^\ast$, respectively. The solid is modelled as an incompressible damped Neo-Hookean solid (hyperelastic solid) with shear modulus $G^\ast$, viscosity $\mu_s^\ast$, and density $\rho_s^\ast$. At relatively low velocities, the flow can be modelled as an incompressible flow, in which case the density and viscosity of the flow, denoted by $\rho_f^\ast$ and $\mu_f^\ast$, respectively, can be treated as constants; and the flow can be modelled by the continuity and momentum equations alone (see \cite{deka2025effect} for formulation). However, to keep the formulation generic for gas-solid systems, we consider the case of a compressible flow. In that case, the energy equation is required to be solved in the flow due to its coupling with the momentum equation. Since the temperature and heat flux is continuous across the solid-fluid interface, the energy equation is also required to be solved inside the solid in addition to the mass and momentum equations. We denote the thermal conductivity of the fluid and the solid media by $\kappa_f^\ast$ and $\kappa_s^\ast$, respectively. The temperature boundary condition is specified at the top moving surface which is maintained at a temperature, $T_w^\ast$, while the bottom surface of the compliant layer can either be kept adiabatic (i.e., zero heat flux) or isothermal at a temperature, $T_{bw}^\ast$. The density, temperature and pressure in the fluid are related by the equation of state (EOS), which in this study is considered to be the ideal gas equation of state. 

We non-dimensionalise length by $L_f^\ast$, velocity by $V_w^\ast$, time by $L_f^\ast/V_w^\ast$, and temperature by $T_w^\ast$. The base flow is considered to be a steady and uni-directional (as indicated diagrammatically in figure \ref{fig-schm}), in which case, there are no variations along the flow ($x$) and spanwise ($z$) directions, i.e., $\partial/\partial x = \partial/\partial z = 0$. This means that the fluid stresses are constants all along the solid-fluid interface implying that the base flow causes only a constant shear deformation in the solid. It is similar to the studies of \cite{ref-patne-2017,deka2025effect}, except in our case, the flow is compressible. The unidirectional flow also implies that the wall-normal component of velocity is zero everywhere, which implies that pressure is constant along the wall-normal ($y$) direction. Therefore, we non-dimensionalise pressure in the fluid with this base flow value $\barp^\ast$. The fluid density is non-dimensionalised by the density scale deduced from the ideal gas EOS, $\rho_w^\ast = \barp^\ast/R T_w^\ast$ evaluated at the top wall. 

\subsection{Governing equations}
\label{subsec:gov_eqs}

The non-dimensionalised compressible Navier-Stokes equations for a calorifically perfect gas can be expressed as,

\begin{equation}
  \frac{\partial \rho}{\partial t} + \frac{\partial v_i}{\partial x_i} = 0 ,
\label{eq:rho_nd}
\end{equation}
\begin{equation}
  \rho\left(\frac{\partial v_i}{\partial t} + v_j \frac{\partial v_i}{\partial x_j} \right) = -\frac{1}{\gamma\Ma^2}\frac{\partial p}{\partial x_i} + \frac{1}{\Re}\frac{\partial }{\partial x_j} \left( \mu\left(\frac{\partial v_i}{\partial x_j} + \frac{\partial v_j}{\partial x_i} \right) - \frac{2\mu}{3} \frac{\partial v_k}{\partial x_k}\delta_{ij} \right) ,
\label{eq:mom_nd}
\end{equation}
\begin{equation}
\begin{aligned}
  \rho\left(\frac{\partial T}{\partial t} + v_j \frac{\partial T}{\partial x_j} \right) =&  -(\gamma - 1)\: p \left(\frac{\partial v_k}{\partial x_k}\right) + \frac{\gamma}{\Re\Pr}\frac{\partial }{\partial x_j} \left( \kappa \frac{\partial T}{\partial x_j}\right) \\
	&\:\: + \frac{\gamma (\gamma - 1) \Ma^2}{\Re}\left( \left( \mu\left(\frac{\partial v_i}{\partial x_j} + \frac{\partial v_j}{\partial x_i} \right) - \frac{2\mu}{3} \frac{\partial v_k}{\partial x_k}\delta_{ij} \right) \frac{\partial v_i}{\partial x_j} \right) ,
\label{eq:ene_nd}
\end{aligned}
\end{equation}
\begin{equation}
 p = \rho T ,
\label{eq:eos_nd}
\end{equation}
where, 
\begin{equation}
\label{eq:nd_defs}
 \Ma = \dfrac{V_w^\ast}{\sqrt{\gamma R T_w^\ast}}\:, \:\: \Re = \dfrac{\rho_w^\ast V_w^\ast L_f^*}{\mu_w^\ast}\:, \:\: \Pr = \dfrac{\gamma R \mu_w^\ast}{(\gamma - 1) \kappa_w^\ast}\:,
\end{equation}
where, $\gamma$ is the ratio of specific heat at constant pressure to constant volume, and $R$ is the universal gas constant. $\mu_w^*$ and $\kappa_w^*$ are the fluid reference values of viscosity and thermal conductivity, respectively, which are considered at the top wall. The non-dimensionalised viscosity and thermal conductivity are expressed as functions of temperature using the Sutherland's relation,
\begin{equation}
 \mu = \kappa = T^{\frac{3}{2}}\left(\frac{1 + C}{T + C}\right),
\end{equation}
where $C = 0.5$ is considered.

The deformation in the solid is modelled by the continuum formulation for a hyperelastic solid material with viscous damping. We first define the deformation tensor,
\begin{equation}
 F_{ij} = \frac{\partial x_i}{\partial X_i} = \delta_{ij} + \frac{\partial u_i}{\partial X_j},
\end{equation}
where, $X_i$ denotes the location of a material point in the undeformed configuration, and $x_i = X_i + u_i$, denotes the location of the material point in the deformed configuration; $u_i$ denoting the deformation. Non-dimensionalising length, time, velocity and temperature in the solid with the same scales as those for the fluid, the mass, momentum and energy equations\footnote{It is assumed that thermoelastic effects are negligible in the solid leading to the energy equation de-coupled from the momentum equation analogous to incompressible fluids.} for an incompressible continuum solid, in non-dimensional form, expressed in the Lagrangian (undeformed) configuration are,
\begin{equation}
  \det(F_{ij}) = 1,
\label{eq:cont_s}
\end{equation}
\begin{equation}
 \frac{\partial^2 {u}_{i}}{\partial t^2} = \frac{\partial P_{ij}}{\partial X_j},
\label{eq:mom_s_lag}
\end{equation}
\begin{equation}
  C_r \frac{\partial T_s}{\partial t}  = -\frac{\partial Q_i}{\partial X_i} ,
\label{eq:ene_s_eul}
\end{equation}
where, $P_{ij} = F^{-1}_{jk} \sigma^{(s)}_{ki}$, is the first Piola-Kirchhoff stress tensor, where $\sigma^{(s)}$ denotes the Cauchy stress tensor in the solid, and, $Q_{i}$ is the heat flux vector expressed in the undeformed configuration. The non-dimensional parameter $C_r = C_s^\ast/(R/(\gamma - 1)$, is the ratio of specific heats of the solid to the fluid (ideal gas). Details of the derivation of the continuum formulation for the solid can be obtained in standard reference books like \cite{ref-holzapfel-2002,ref-gonzalez-2008,ref-jog-2015}. For an incompressible Neo-Hookean solid with viscous damping the Cauchy stress tensor in non-dimensional form can be written as,
\begin{equation}
 \sigma^{(s)}_{ij} = - \frac{p_s \delta_{ij}}{\gamma \Ma^2}  + \frac{b_{ij}}{\lambda} + \frac{\mu_r}{\rho_r \Re} \left(l_{ij} + l_{ji}\right), 
\end{equation}
where, $b_{ij}$ and $l_{ij}$ are the left Cauchy-Green tensor and the velocity gradient tensor, respectively, expressed in terms of the deformation gradient tensor as,
\begin{equation}
 b_{ij} = F_{ik} F_{jk} \:, \:\: l_{ij} = \frac{\partial F_{ik}}{\partial t} F^{-1}_{kj},
\end{equation}
and, $p_s$ is the incompressible pressure in the solid, obtained after non-dimensionalising by the scale $\rho_s^\ast R T_w^\ast$. 
The non-dimensional parameters, $\rho_r = \rho_s^\ast/\rho_w^\ast$ and $\mu_r = \mu_s^\ast/\mu_w^\ast$ are the solid-to-fluid ratios of density and viscosity, respectively, while the elasticity parameter, $\lambda$, is defined as,
\begin{equation}
 \lambda = \frac{\rho_s^\ast V_w^{\ast 2}}{G^\ast} = \frac{V_w^{\ast 2}}{V_s^{\ast 2}},
\label{eq:lambda_def}
\end{equation}
where, $V_s^\ast = \sqrt{G^\ast/\rho_s^\ast}$, represents shear wave speed in the solid. It is easy to check that the elasticity parameter, $\lambda$, can be related to the other elasticity parameter $\Gamma = \mu_w^\ast V_w^\ast/ L_f^\ast G^\ast$ chosen in previous works \cite{ref-kumaran-2021,ref-patne-2017,ref-patne-2019,deka2025effect} as, $\lambda = \rho_r \Gamma \Re$. While the formulation of the solid equations are similar to the studies of \cite{ref-patne-2017,ref-patne-2019}, the non-dimensionalisation adopted here is marginally different owing to its better suitability for high solid-to-fluid density ratio cases. Finally, the non-dimensional heat flux vector is closed using the Fourier law of heat conduction \cite{ref-holzapfel-2002}, 
\begin{equation}
 Q_i = -\frac{\gamma \kappa_r}{\rho_r \Re\Pr} (F^T F)^{-1}_{ij}\frac{\partial T_s}{\partial X_j},
\end{equation}
where, $\kappa_r = \kappa_s^\ast/\kappa_w^\ast$, is the ratio of thermal conductivity in the solid to the fluid reference value. 

For our system, the governing equations are solved with the top moving wall having a no-slip ($v_x = 1$, $v_y = 0$) and isothermal ($T = 1$) boundary condition, while the bottom surface of the compliant solid layer has a zero deformation ($u_x = u_y = 0$) and either isothermal ($T_s = T_{bw}$) or adiabatic ($\partial T_s/\partial y = 0$) boundary condition. At the solid-fluid interface, the velocity, stress, temperature and heat flux are required to be continuous, which in non-dimensional form can be written as,
\begin{equation}
\begin{aligned}
 v_i &= \frac{\partial u_i}{\partial t},\\
 -\frac{{\Re}}{\gamma\Ma^2}p\delta_{ij} + {\mu}\left(\frac{\partial v_i}{\partial x_j} + \frac{\partial v_j}{\partial x_i} - \frac{2}{3}\frac{\partial v_k}{\partial x_k}\delta_{ij}\right) &= -\frac{\rho_r \Re}{\gamma\Ma^2}p_s\delta_{ij} + \frac{\rho_r\Re}{\lambda}b_{ij} + \mu_r(l_{ij} + l_{ji}),\\
 T &= T_s,\\
 \kappa \frac{\partial T}{\partial x_i} &= \kappa_r \frac{\partial T_s}{\partial x_i}. 
\end{aligned}
\label{eq:bc_full}
\end{equation}

To study the stability of the system, we consider the steady unidirectional base state and add two-dimensional small amplitude perturbations to the entire system, i.e,
\begin{equation}
\label{eq:ansatz1}
  \mathbf{q} = \bar{\mathbf{q}} + \epsilon\mathbf{q}^{\prime} 
\end{equation}
where, $\mathbf{q} = [ \rho \:\: v_x \:\: v_y \:\: T \:\: u_x\:\: v_y\:\: p_s\:\: T_s]^{T}$, represents the solution vector with the components as fluid density, fluid $x$-velocity, fluid $y$-velocity, fluid temperature, solid $x$-displacement, solid $y$-displacement, solid pressure and solid temperature, respectively. The quantities $\bar{(.)}$ represent the steady one-dimensional base state field, and the quantities $(.)^\prime$ represent the perturbation field with $\epsilon \ll 1$ denoting small amplitude perturbations. 

In this study, the governing equations in the solid are expressed in the base-state deformed configuration which is referred to as the `L3' formulation \cite{ref-patne-2017} (see figure \ref{fig-config}). If $X_i$ denotes the location of a material point in the original undeformed state of the solid, $\barx_i$ denotes the location in the base-state deformed configuration and $x_i$ denotes the final deformed configuration (after addition of perturbations), we can write,
\begin{equation}
\begin{aligned}
 \barx =& X + \barus(Y) \:,\:\: \bary = Y\:, \\
 x =& \barx + \epsilon \pus(\barx,\bary) \:,\:\: y = \bary + \epsilon \pvs(\barx,\bary)\:,
\end{aligned}
\label{eq:solid_frame}
\end{equation}
for the case of small amplitude perturbations ($\epsilon \ll 1$) imposed on a steady sheared solid in the base state. 

\subsection{Base state}

The base state deformation in the solid, obtained by solving the $x$-momentum equation, which reduces to $dP_{XY}/dY = 0$ for a steady unidirectional flow, is (see \cite{ref-kumaran-2021} for detailed derivation),
\begin{equation}
 \barus = \frac{\lambda}{\rho_r \Re} \left(\bar{\mu} \frac{d\baru}{dy}\right)\Bigg|_{\bary = 0} (\bary + L_r) ,
\label{eq:base_state_ux}
\end{equation}
while $\barvs = 0$, due to the incompressibility constraint. The constants of integration are obtained by satisfying the zero deformation condition at the bottom wall ($\bary = -L_r$, where $L_r = L_s^\ast/L_f^\ast$) and by continuity of the shear stress at the interface ($\bary = 0$). Note that the base state solid deformation becomes identical to the one obtained in \cite{deka2025effect} if the fluid shear stress at the interface is set to unity, which holds true if the flow is incompressible. For a compressible flow, the fluid shear stress at the interface can to be obtained by solution of the base flow equations. For a unidirectional compressible flow, the continuity equation is trivially satisfied and the normal momentum equation reduces to zero pressure gradient condition, while the tangential momentum equation and energy equations simplify to,
\begin{equation}
 \frac{d}{dy}\left(\barmu \frac{d\baru}{dy} \right) = 0 ,
\label{eq:base_umom}
\end{equation}
\begin{equation}
 \frac{d}{dy}\left(\barkappa \frac{d\barT}{dy} \right) + (\gamma - 1) \Ma^2 \Pr \barmu \left(\frac{d\baru}{dy}\right)^2 = 0 .
\label{eq:base_ene}
\end{equation}
The above equations are solved numerically with appropriate boundary conditions. Since the base-state deformation of the solid is steady, the velocity continuity at the interface implies that the fluid velocity at the solid-fluid interface has to be zero. The interface condition for temperature can be deduced by imposing the continuity of temperature and heat flux at the interface. For that, we, first solve for the temperature field in the solid. For a steady-unidirectional deformation case, the energy equation in the solid simplifies to $d^2 \barT_s/d\bary^2 = 0$, which leads to a linear solution. 
\begin{itemize}
\item If the bottom surface of the solid layer is assumed to be adiabatic ($d \barT_s /d\bary|_{y=-L_r} = 0$), then, it implies that the temperature in the solid has to be a constant, and by heat flux continuity it implies that at the interface,
\begin{equation}
\frac{d\barT}{dy}\Big|_{\bary = 0} = 0,
\label{eq:temp_adia_bw}
\end{equation}
becomes the appropriate condition for the fluid temperature at the interface.
\item If the bottom surface of the solid layer is assumed to be isothermal ($\barT_s = T_{bw}$), then the solution of temperature in the solid becomes,
\begin{equation}
 \barT_s = T_{bw} + \left(\frac{\bar{\kappa}}{\kappa_r} \frac{d\barT}{dy}\Bigg|_{\bary = 0}\right) (\bary + L_r) ,
\end{equation}
therefore, by temperature continuity, the interface condition for temperature becomes a Robin boundary condition,
\begin{equation}
\barT|_{\bary = 0} = T_{bw} + \left(\frac{\bar{\kappa}}{\kappa_r} \frac{d\barT}{dy}\Bigg|_{\bary = 0}\right) L_r.
\label{eq:temp_iso_bw}
\end{equation}
\end{itemize}
To solve the perturbation equations, the interface conditions for the perturbations have to be imposed at the correct interface location after imposition of the perturbations. For this, it has to be noted that the equations for the solid are posed in the base-state deformed configuration while the fluid equations are expressed in the deformed configuration (Eulerian), therefore, the boundary conditions have to be consistently interpolated as explained in the previous works of \cite{ref-patne-2017,deka2025effect}.

\subsection{Stability formulation}

In this work, we study the stability of the system to two-dimensional normal-mode perturbations of the form,
\begin{equation}
 \mathbf{q}^\prime = \hat{\mathbf{q}}\exp\left(\imath\kx(x - ct)\right),
\end{equation}
which is substituted into the ansatz shown in Eq. \ref{eq:ansatz1} and the solution then substituted into the governing equations, Eqs. \ref{eq:rho_nd}-\ref{eq:eos_nd} and Eqs. \ref{eq:cont_s}-\ref{eq:ene_s_eul} and boundary conditions, Eqs. \ref{eq:bc_full}. The $O(\epsilon)$ equations and boundary conditions obtained after the above substitution are shown in appendix \ref{secapp:modal_eqs}. The temporal stability of the system is studied, therefore, the wave-number $\kx$ is real and the wave-speed $c$ is complex is general. The normal mode equations are cast into a quadratic eigenvalue problem of the form,
\begin{equation}
 \left(\mathcal{L}_0 + c \mathcal{L}_1 + c^2 \mathcal{L}_2\right) \mathbf{\hat{q}} = 0, 
\label{eq:stab_prob}
\end{equation}
which is solved numerically using Chebyshev-Pseudospectral method (see \cite{ref-deka-2023} for solver details). The eigenfunctions in the solid and fluid regions are expanded individually using $N$-th order Chebyshev polynomials, where $N$ is the number of collocation points. The resulting matrix-eigenvalue equations are solved using the function \texttt{polyeig} in MATLAB. From numerical tests, it is observed that $N=100$ gives eigenvalues accurate upto 7th decimal place, and hence is chosen for all calculations in this study.

\begin{figure}
  \begin{center}
  \subfigure[Streamwise velocity]{{\label{fig-base_u}}\includegraphics[width=2.8in]{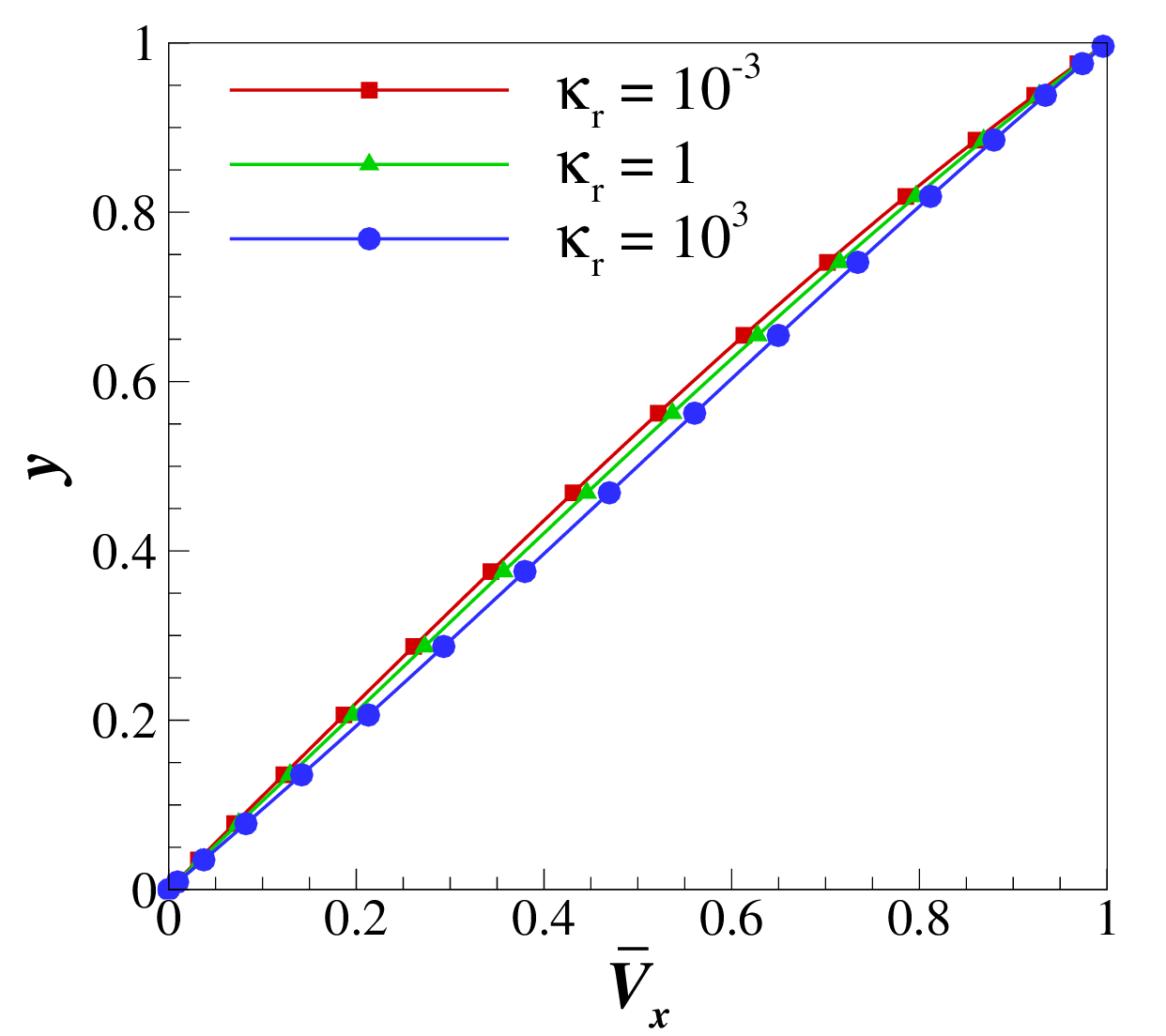}}
  \subfigure[Temperature]{{\label{fig-base_T}}\includegraphics[width=2.8in]{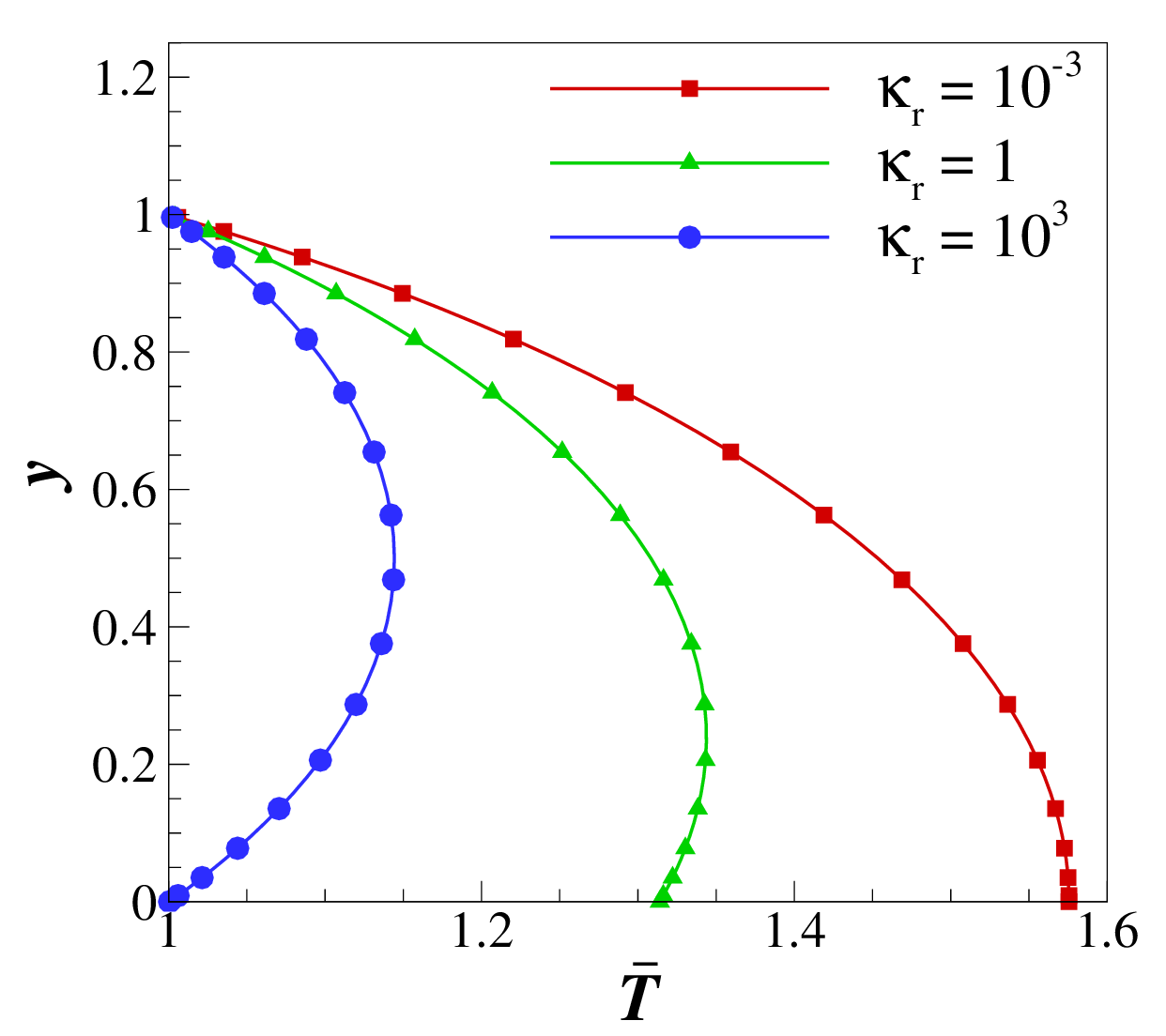}}
  \caption{Base flow profiles for (a) streamwise velocity, and (b) temperature, at $\Ma = 2$ for an isothermal bottom wall with $T_{bw} = 1$, for different values of thermal conductivity ratio, $\kappa_r$.} 
    \label{fig-base_kappar_var}
  \end{center}
\end{figure}

\section{Numerical results}
\label{sec:results}

In this section, the numerical solutions of the stability problem are discussed. As evident from the mathematical formulation of the problem, there are a number of non-dimensional parameters that could potentially affect the stability of the system. In this work, we primarily investigate the effect of the parameters, $\Ma$, $\lambda$, $\kappa_r$, $\mu_r$ and, $T_{bw}$. For all calculations, we consider the solid and fluid layers to be of equal thickness, i.e., $L_r = 1$, the solid-to-fluid density ratio as, $\rho_r = 1000$, which is characteristic of gas-solid systems, and parameters, $\Re = 100$, $\gamma = 1.4$ and $\Pr = 0.71$ are chosen, unless stated otherwise.

\subsection{Base flow}
\label{subsec:base_flow_sols}

The base flow equations, Eqs. \ref{eq:base_umom}-\ref{eq:base_ene}, are solved numerically with $\baru = 1$ and $\barT = 1$ at the top wall, and $\baru = 0$ and, either Eq. \ref{eq:temp_adia_bw} for an adiabatic bottom wall, or Eq. \ref{eq:temp_iso_bw} for an isothermal bottom wall. For the adiabatic case, the base flow profiles obtained are same as the ones obtained for a rigid wall compressible Couette flow case and have been extensively discussed in the previous works of \cite{ref-duck-94,ref-hu-98}. For the isothermal bottom wall case, the interface condition at $\bary = 0$ is a mixed (Robin) boundary condition that depends on the conductivity ratio and bottom wall temperature. Figure \ref{fig-base_kappar_var} shows the streamwise velocity and temperature variation in the base flow computed at $\Ma = 2$ and $T_{bw} = 1$. As evident from the figures, the base temperature profiles show appreciable variation upon changing conductivity ratio, but the streamwise velocity profiles remain relatively unchanged. At low conductivity ratio ($\kappa_r = 10^{-3}$) the temperature profile resembles close to an adiabatic bottom wall case which is because the solid behaves as an insulating layer. Similarly, for a highly conducting solid ($\kappa_r = 10^3$), the temperature at the interface is approximately equal to the bottom wall value. Therefore, increasing conductivity ratio changes the interface temperature from the adiabatic wall temperature value ($T_{ad}$) to the imposed value at the bottom wall of the solid layer ($T_{bw}$). The change in the base temperature solution upon changing conductivity ratio affects the stability of the flow. 

\subsection{Stability solutions}

The stability of the given system is determined by numerical computation of the eigenvalue problem, Eq. \ref{eq:stab_prob}. The numerical solver developed to compute the eigenvalues has been validated by comparing the low-$\Ma$-finite-$\lambda$ results against the incompressible results of \cite{deka2025effect}, and by comparing the low-$\lambda$-finite-$\Ma$ results against rigid wall compressible Couette flow results of \cite{ref-duck-94,ref-hu-98}. 

\begin{figure}
\begin{center}
\begin{minipage}{.45\textwidth}
  \includegraphics[width=3in]{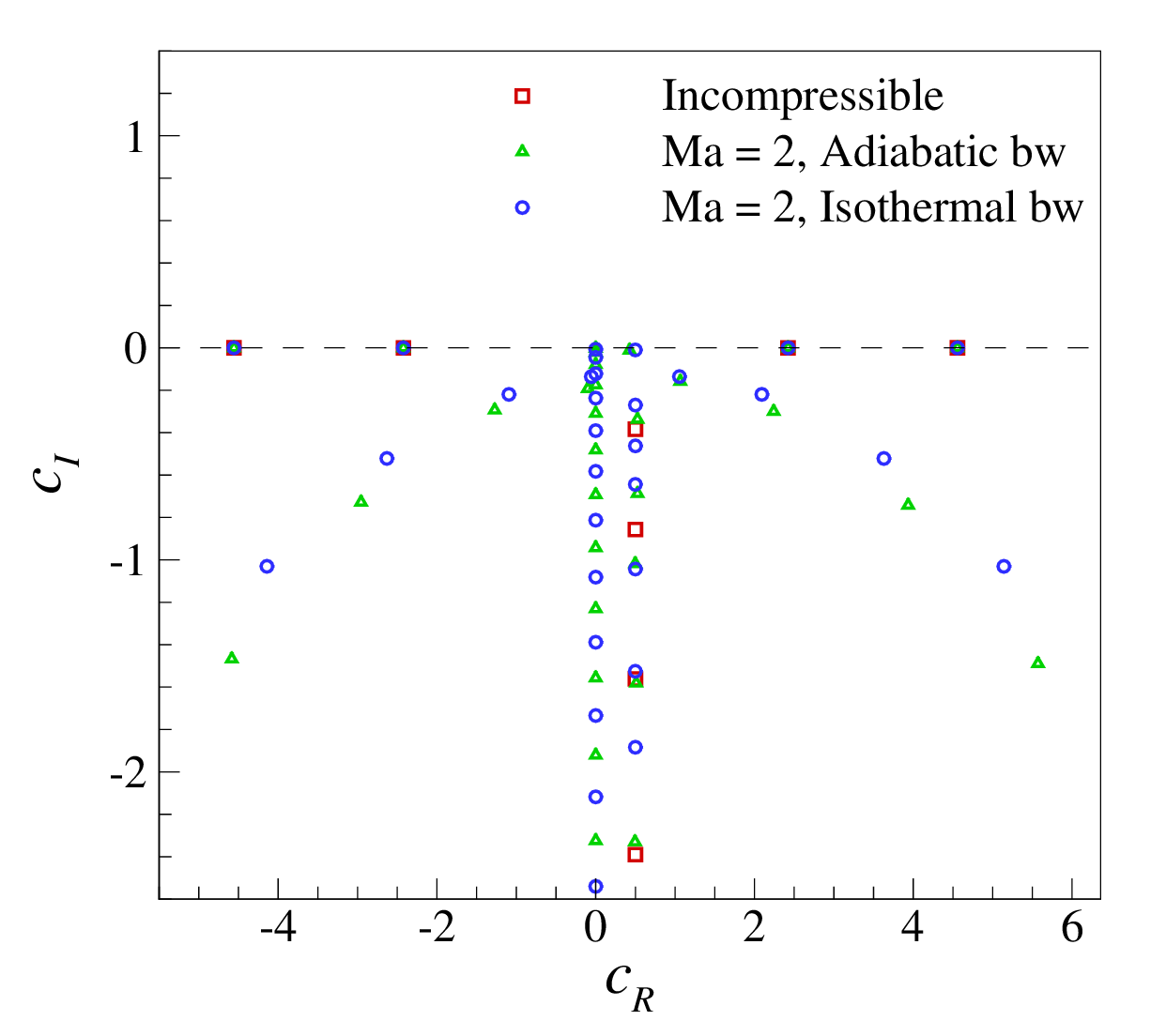}
  \caption{Eigenspectrum at $\kx = 1$ for incompressible, $\Ma = 2$ with adiabatic bottom wall, and $\Ma=2$ with isothermal bottom wall ($T_{bw} = 1$), with $\kappa_r = 100$ for the compressible cases. For all cases, $\lambda = 1$, $\Re = 100$, $\mu_r = 0$.}
  \label{fig-spectra-1}
\end{minipage}
\quad
\begin{minipage}{.45\textwidth}
  \includegraphics[width=3in]{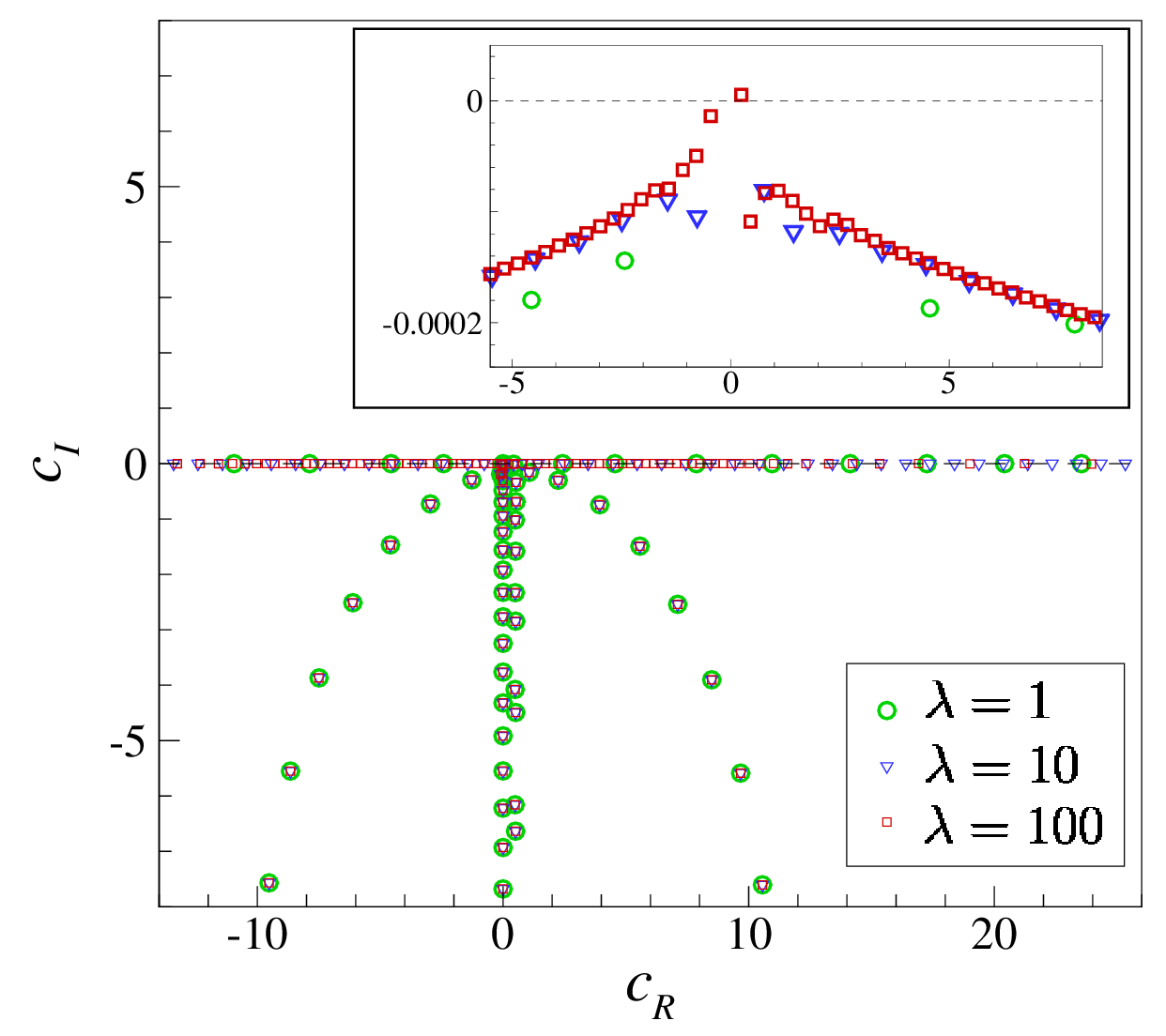}
  \caption{Eigenspectrum at $\kx=1$ computed for different values of flexibility parameter $\lambda$ for the adiabatic bottom wall case. Inset shows the zoomed-in eigenspectra with an unstable mode ($\cI>0$) at $\lambda = 100$. For all cases, $\Ma = 2$, $\Re = 100$, $\mu_r = 0$, $\kappa_r = 100$.}
  \label{fig-spectra-2}
\end{minipage}
\end{center}
\end{figure}

For the case of a compressible flow past a viscoelastic compliant solid, the eigenspectra can be qualitatively categorised into four distinct classes of modes. These can be shown in the eigenspectra plotted in figure \ref{fig-spectra-1}, which is computed at $\kx = 1$, for $\lambda = 1$, $\mu_r = 0$, for the incompressible case\footnote{The formulation for the incompressible case is followed from \cite{deka2025effect}.}, and two compressible cases at $\Ma = 2$, one with an adiabatic bottom boundary and one with an isothermal bottom boundary with $T_{bw} = 1$. 
The first class of modes are referred to as the `incompressible flow' modes. These modes exist in the incompressible flow past rigid walls which are, here, modified by flow compressibility and wall flexibility. These are characterised with the wavespeed $\cR \approx 0.5$ in figure \ref{fig-spectra-1} and are evidently present for all three cases.
The second are the `compressible flow' modes which arise due to flow compressibility (finite $\Ma$) and are modified by wall flexibility. These modes are often referred to as `higher' modes in compressible stability literature \cite{ref-lees_lin-46,ref-mack-84,ref-duck-94} and can be shown in figure \ref{fig-spectra-1} as the two families that form the `umbrella' structure in the compressible flow cases. The left branch modes, $\cR<0$, constitute modes II, IV, \ldots, and the right branch modes, $\cR>1$, constitute modes I, III, \ldots, of a rigid wall compressible Couette flow (nomenclature followed from \cite{ref-duck-94,ref-deka-2023-channel}). At $\Re = 100$, these modes are relatively stable and typically become unstable in rigid wall cases when $\Re \approx 10^4$ or higher.
The third class of modes are the `solid' modes that exist due to elastic stresses in the solid and get modified due to interaction with the flow. These modes also show as two families, one with $\cR < 0$ and the other with $\cR> 0$ and are characterised by $\cI$ very close to $0$ in figure \ref{fig-spectra-1}. These modes are present in all three cases and are considerably more unstable (less negative values of $\cI$) compared to the other families. 
The final class of modes are the thermal modes in the solid which are the eigensolutions of the energy equation of the solid, characterised by $\cR \approx 0$ and $\cI<0$. Numerical studies have shown that these modes are always stable and hence are not discussed further in this work. 

\begin{figure}
\begin{center}
\begin{minipage}{.45\textwidth}
  \includegraphics[width=3in]{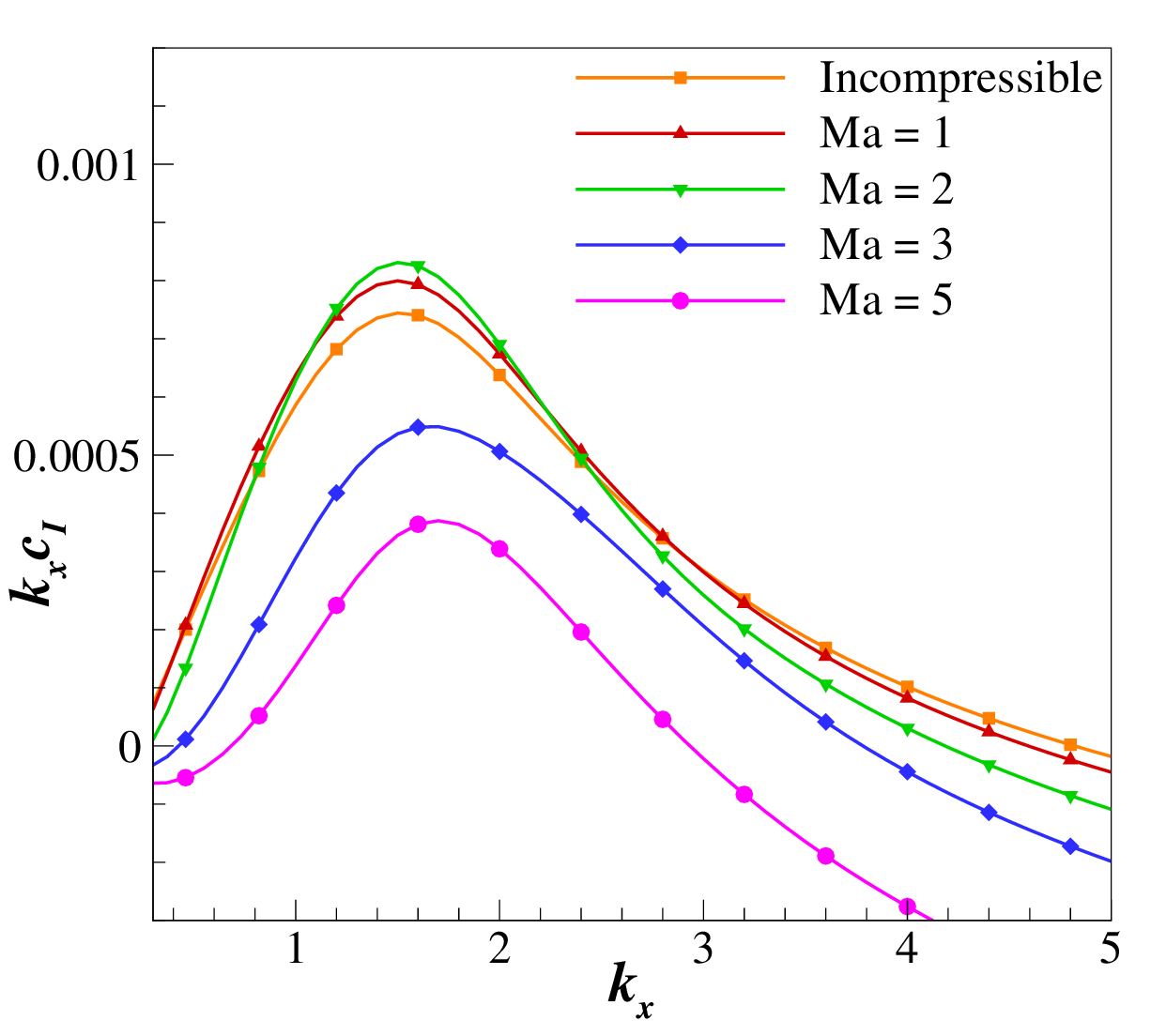}
  \caption{Variation of growth rate ($\kx\cI$) with streamwise wave-number ($\kx$) for adiabatic bottom wall case at different values of $\Ma$. For all cases, $\Re = 100$, $\lambda = 1000$, $\mu_r = 0$, $\kappa_r = 100$.}
  \label{fig-disp_rel_ma}
\end{minipage}
\quad
\begin{minipage}{.45\textwidth}
  \includegraphics[width=3in]{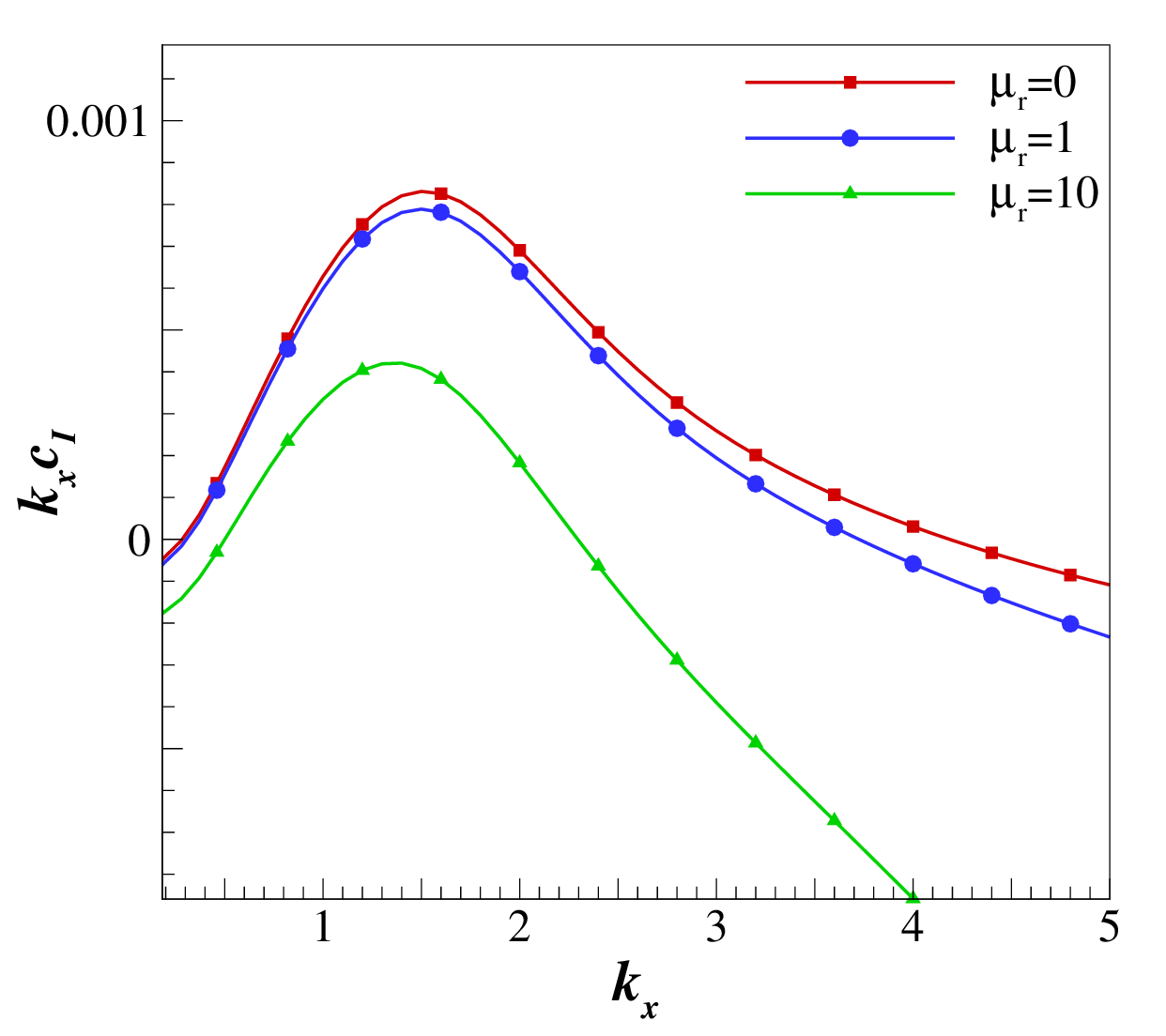}
  \caption{Variation of growth rate ($\kx\cI$) with streamwise wave-number ($\kx$) for adiabatic bottom wall case at different values of $\mu_r$. For all cases, $\Ma = 2$, $\Re = 100$, $\lambda = 1000$, $\kappa_r = 100$.}
  \label{fig-disp_rel_mur}
\end{minipage}
\end{center}
\end{figure}

Among the four classes of modes, the most dominant instability has been numerically observed to belong to the third category, i.e., the elastic modes of the solid. These modes are effectively shear waves of the elastic solid and have been previously studied in the context of incompressible flows \cite{kumaran1995effect}. The elastic modes can exist only for finite wall elasticity, characterised here by the parameter $\lambda$. $\lambda$, as defined in Eq. \ref{eq:lambda_def}, is the square of the ratio of the flow speed to the shear-wave speed, therefore, in the asymptotic limit of $\lambda\rightarrow 0$, the wavespeed, $\cR \rightarrow \pm\infty$, for the two families of the elastic modes. At finite values of $\lambda$, the elastic modes become unstable due to energy exchange with the flow across the interface. Figure \ref{fig-spectra-2} shows the eigenspectra at $\Ma = 2$ and $\Re=100$, with increasing value of $\lambda$ from $1$ to $100$. As $\lambda$ is increased, the magnitude of $\cR$ decreases for the solid modes and, from the figure inset it can be seen that, the first elastic mode with $\cR>0$ becomes unstable ($\cI>0$) at $\lambda = 100$. It is to be noted that typical materials used for compliant coatings, like elastomers, have quite low values of shear wave-speeds, $\sim 10$ m/s, whereas the flow speed for external applications is at least one order of magnitude larger, therefore, values of $\lambda$ of the order of $100-1000$ are practically relevant, and in this regime, the elastic modes are critical in determining the stability of the system. The other important observation from figure \ref{fig-spectra-2} is that the ``fluid'' modes, i.e., the first and second class of modes do not show any appreciable change in the eigenvalues upon increasing $\lambda$, i.e., wall flexibility has a negligible effect on the eigenmodes of the flow. Therefore, the subsequent analysis in this work will focus only on the elastic modes, since the destabilisation of the fluid modes will require much higher operational Reynolds numbers, whereas at $\Re$ as low as $100$, the elastic modes clearly destabilise at practically relevant values of $\lambda$.  

\begin{figure}
\begin{center}
\begin{minipage}{.45\textwidth}
  \includegraphics[width=3in]{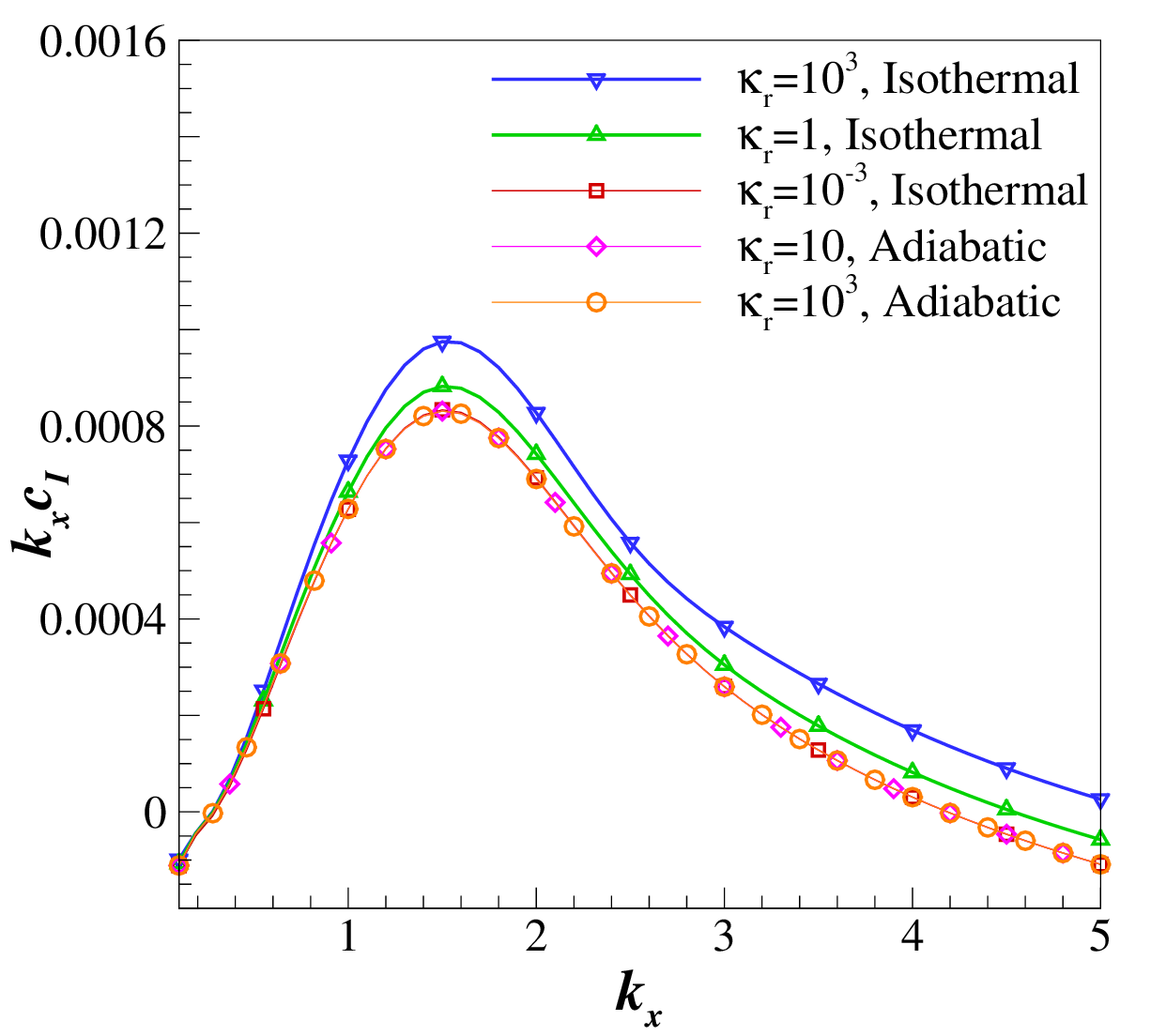}
  \caption{Variation of growth rate ($\kx\cI$) with streamwise wave-number ($\kx$) for adiabatic and isothermal bottom wall cases at different values of $\kappa_r$. For all cases, $\Ma = 2$, $\Re = 100$, $\lambda = 1000$, $\mu_r = 0$, and $T_{bw}=1$ (for the isothermal bottom wall cases).}
  \label{fig-disp_rel_kappar}
\end{minipage}
\quad
\begin{minipage}{.45\textwidth}
  \includegraphics[width=3in]{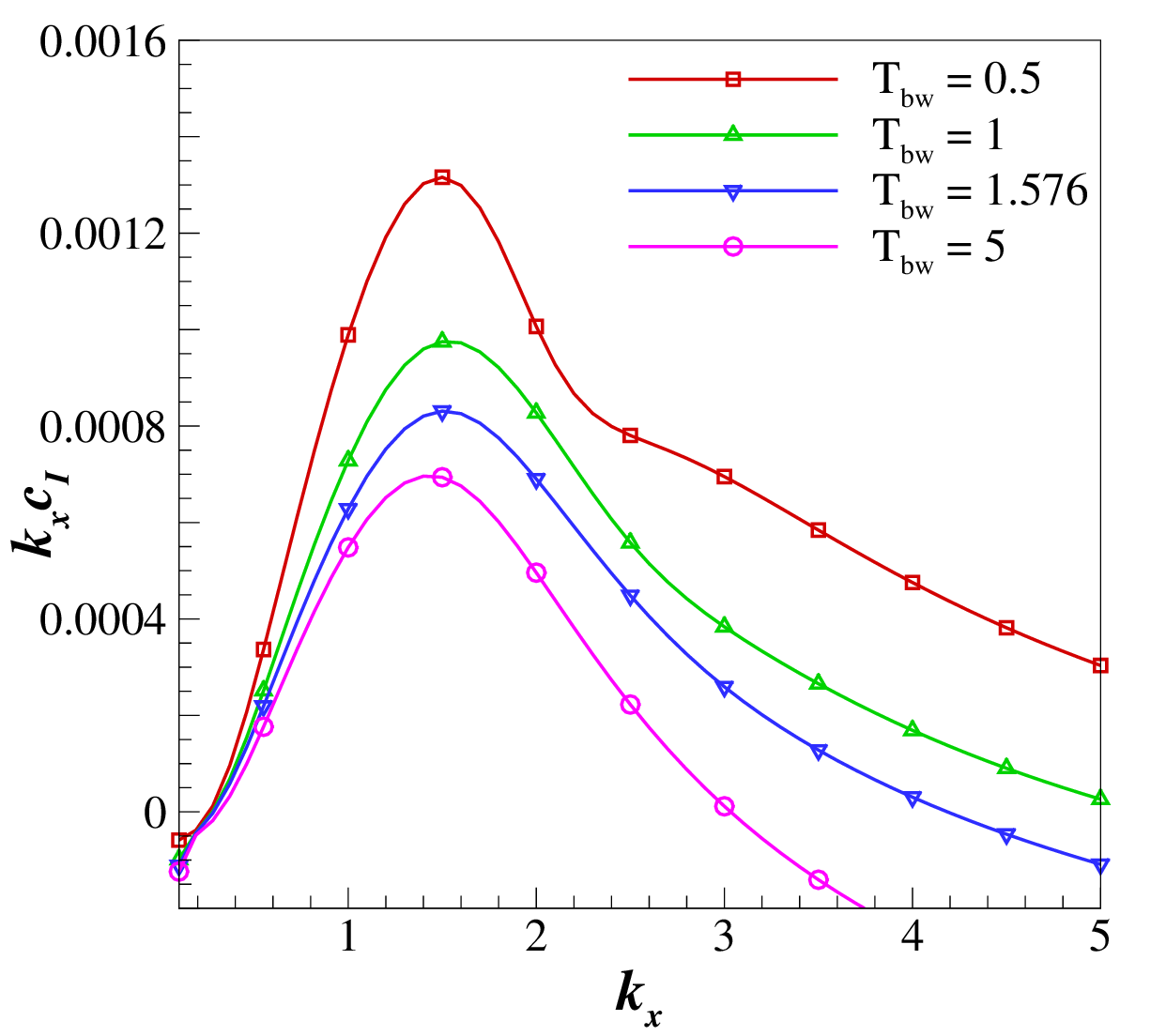}
  \caption{Variation of growth rate ($\kx\cI$) with streamwise wave-number ($\kx$) for the isothermal bottom wall cases at different values of bottom wall temperature. For all cases, $\Ma = 2$, $\Re = 100$, $\lambda = 1000$, $\mu_r = 0$, and $\kappa_r = 1000$.}
  \label{fig-disp_rel_Tbw}
\end{minipage}
\end{center}
\end{figure}

We study the effect of the flow compressibility on the stability of the elastic modes. Figure \ref{fig-disp_rel_ma} shows the variation of the growth rate ($\kx\cI$) of the first elastic mode (which is also the most dominant mode) with wave-number ($\kx$) for the incompressible case and compressible cases with increasing value of Ma. It is observed from the figure that upto $\Ma \approx 2$, increasing compressibility increases the maximum growth rate of the instability, and after that has a stabilizing influence. This non-monotonic dependence on flow compressibility is in direct contrast to the work of \cite{deka2025linear}, where stability of a compressible channel flow was studied past a compliant wall modelled as a spring-backed-plate. In that study, it was shown that increasing flow compressibility had a purely stabilizing effect on the travelling wave flutter (TWF) mode, which is equivalent to the elastic modes of our system. The non-monotonic dependence of $\Ma$ on the stability of the elastic modes can be explained using an asymptotic analysis, which is discussed in the next section. In all studies so far, the solid is assumed to be non-dissipative, i.e., $\mu_r = 0$. In figure \ref{fig-disp_rel_mur}, the growth rate of the first elastic mode has been studied by increasing $\mu_r$ from $0$ to finite values. It is seen that due to its dissipative nature, increasing solid viscosity has a stabilizing effect on the elastic modes, which is also consistent with what is observed for incompressible flow cases \cite{deka2025effect}. Similarly, increasing flow viscosity (decreasing $\Re$) also has a dissipative effect on the elastic modes and hence leads to stability, but has not been shown here for the sake of brevity.
While the results shown in figures \ref{fig-disp_rel_ma} and \ref{fig-disp_rel_mur} are for the case of an adiabatic bottom wall, the same observations hold qualitatively for the isothermal bottom wall case. 

For compressible flow, another important parameter that influences the stability of the elastic modes is the solid-to-fluid conductivity ratio ($\kappa_r$). Figure \ref{fig-disp_rel_kappar} shows the growth rates ($\kx\cI$) as a function of wave-number ($\kx$) for the first elastic mode for both the isothermal and adiabatic bottom wall cases at different values of conductivity ratio. For the isothermal bottom wall case, the figure shows that increasing $\kappa_r$ has a clear destabilizing influence on the elastic mode. For the adiabatic bottom wall case, $\kappa_r$ has no apparent influence on the stability of the mode. Additionally the growth rates of the adiabatic bottom wall case match well with the isothermal bottom wall case at $\kappa_r = 10^{-3}$. It is discussed in section \ref{subsec:base_flow_sols} that changing conductivity ratio significantly alters the base flow temperature profiles for the isothermal bottom wall case. It is also shown that for very low values of $\kappa_r$, the temperature profile becomes identical to the adiabatic bottom wall case. Therefore, it can be asserted that the effect of $\kappa_r$ on the stability of the mode is through the change in the base flow temperature profile. This is why changing $\kappa_r$ has almost no influence on stability for the adiabatic bottom wall case. The effect of bottom wall temperature for the isothermal bottom wall case is shown in figure \ref{fig-disp_rel_Tbw}, where the growth rate of the first elastic mode is shown for different bottom wall temperatures. From the figure it is evident that increasing the bottom wall temperature has a stabilizing effect on the elastic mode. In other words, wall cooling has a destabilizing effect on the elastic mode, which is similar in nature to the mode II instability of a compressible boundary layer \cite{ref-mack-84}.

\section{Asymptotic analysis}
\label{sec:asymptotic}

In this section, we present an asymptotic analysis for the case of a flow past a compliant viscoelastic solid in the high solid-to-fluid density ratio limit. For typical gas-solid systems, $\rho_r \gg 1$ and therefore, we define a small parameter, $\epsilon = (\rho_r\Re)^{-1}$, assuming $\Re \geq 1$. The normal mode solution can be expressed as a regular perturbation expansion in $\epsilon$,
\begin{equation}
 \mathbf{\hat{q}} = \mathbf{\hat{q}}^{(0)} + \epsilon \mathbf{\hat{q}}^{(1)} + O(\epsilon^2),\:\: c = \czero + \epsilon\cone + O(\epsilon^2), 
\end{equation}
which is substituted into the governing equations, Eqs. \ref{rho_mode}-\ref{T_mode_solid}, and boundary conditions, Eqs. \ref{eq:bc1_mode}-\ref{eq:ic6_mode}, in appendix \ref{secapp:modal_eqs}, to obtain the ordered set of equations. Since $\epsilon$ does not appear in the fluid equations (Eqs. \ref{rho_mode}-\ref{p_mode}, appendix \ref{secapp:modal_eqs}), they remain the same at each order of $\epsilon$. The small parameter appears in the solid equations  (Eqs. \ref{rho_mode_solid}-\ref{v_mode_solid}, appendix \ref{secapp:modal_eqs}) through the viscous term and also through the base state deformation, which is shown in Eq. \ref{eq:base_state_ux}, and can be written as, $\barus = \epsilon \lambda \barsigmatw (\bary + L_r)$, where, $\barsigmatw = (\barmu d\baru/dy)|_{y=0}$. For the interface conditions, $\epsilon$ appears only in the normal and tangential stress balance conditions (Eqs. \ref{eq:ic3_mode} and \ref{eq:ic4_mode}, appendix \ref{secapp:modal_eqs}), which at the leading order, reduce to zero stress condition for the solid,
\begin{equation}
 -\frac{\tpszero}{\gamma\Ma^2} + \frac{2}{\lambda} \frac{d\tvszero}{d\bary} = 0,
\label{eq:ic3_mode_incomp_highrhor-zero} 
\end{equation}
\begin{equation}
\frac{d\tuszero}{d\bary} + \imath\kx\tvszero = 0.
\label{eq:ic4_mode_incomp_highrhor-zero}
\end{equation}  
The leading order mass and momentum equations in the solid (Eqs. \ref{eq:rho_mode_solid_highrhor-zero}-\ref{eq:v_mode_solid_highrhor-zero}, appendix \ref{secapp:asymp_eqs}) along with the above zero-stress conditions form a closed system. The eigenmodes of that system are the elastic modes of the solid. The flow then affects these eigenmodes through the boundary interaction at the next order. Thus, it can be seen that in the high $\rho_r\Re$ limit, the fluid-solid interaction de-couples at each asymptotic order.
The coupling happens as follows. The interface deformation obtained after solving the leading order equations in the solid, provide the values for the velocity perturbations at the interface through the velocity continuity conditions (Eqs. \ref{eq:ic1_mode_incomp_highrhor-zero}-\ref{eq:ic2_mode_incomp_highrhor-zero}, appendix \ref{secapp:asymp_eqs}). For an incompressible flow, the continuity and momentum equations in the fluid, at the leading order, can now be standalone solved with these velocity conditions at the interface. For a compressible flow, the flow-thermodynamics coupling implies that a temperature condition at the interface is additionally required to solve the fluid equations at the leading order. 
But the temperature interface conditions (Eqs. \ref{eq:ic5_mode} and \ref{eq:ic6_mode}, appendix \ref{secapp:modal_eqs}) do not directly de-couple as $\epsilon$ does not appear in these conditions. However, it can be shown that $\tTzero + (d\barT/d\bary)\tvszero = 0$ is a good enough approximation for the temperature interface condition at the leading order (see appendix \ref{secapp:ene_solid} for the formal analysis). Therefore, with this condition, we can solve the fluid equations at the leading order for a general compressible flow. Once, the leading order eigenfunctions in the fluid are obtained, they provide the boundary conditions for the first-correction to the solid mass and momentum equations, through the first-correction to the stress continuity conditions (Eqs. \ref{eq:ic3_mode} and \ref{eq:ic4_mode}, appendix \ref{secapp:modal_eqs}), which read as,
\begin{equation}
 -\frac{\tpsone}{\gamma\Ma^2} + \frac{2}{\lambda}\left(\frac{d\tvsone}{d\bary} + \imath\kx\lambda\barsigmatw\tvszero \right) - {2\imath\kx \czero \mu_r}\frac{d\tvszero}{d\bary} = -\frac{\Re\tpzero}{\gamma\Ma^2} + {2\barmu}\frac{d\tvzero}{dy} - \frac{2\barmu}{3}\left(\imath\kx\tuzero + \frac{d\tvzero}{dy} \right),
\label{eq:ic3_mode_highrhor-first} 
\end{equation}
\begin{equation}
\frac{1}{\lambda} \left(\frac{d\tusone}{d\bary} + \imath\kx\tvsone\right) = {\barmu} \left(\frac{d\tuzero}{dy} + \imath\kx\tvzero \right) + \frac{d\mu}{dT}\Bigg|_{\barT} \frac{d\baru}{dy} \tTzero + \frac{d}{dy}\left(\barmu\frac{d\baru}{dy}\right) \tvszero ,
\label{eq:ic4_mode_highrhor-first}
\end{equation}
Thus, the $O(\epsilon)$ mass and momentum equations in the solid can now be solved with the above conditions to obtain the first-correction to the eigenfunctions in the solid. Subsequently, the velocity continuity conditions can be used to obtain the interface values of the first-correction to the velocity eigenfunctions. The first-correction to the fluid equations can then be solved to obtain the fluid eigenfunctions and the process continued to obtain higher order corrections. 

The main objective of our analysis is to obtain an approximation for the eigenvalue corresponding to the most dominant instability, which are the elastic modes. We first combine the mass and momentum equations in the solid into a single evolution equation for the normal deformation. To achieve this, we first take the derivative of Eq. \ref{u_mode_solid} with respect to $y$ and subtract from $\imath\kx$ times Eq. \ref{v_mode_solid} to eliminate pressure, then substituting $\tus = {\imath}/{\kx}({d\tvs}/{d\bary})$ from Eq. \ref{rho_mode_solid} (same procedure adopted as the one in deriving Orr-Sommerfeld equation \cite{ref-schmid-98}). The resulting equation expressed upto $O(\epsilon)$ terms reads as,
 \begin{equation}
\left(D^4 - (2 - \lambda c^2) \kxsq D^2 + (1 - \lambda c^2)\kx^4 \right) \tvs + \epsilon \left(-\lambda \imath \kx c \mu_r \left( D^2 - \kxsq \right)^2 + 2 \lambda \imath \kx \barsigmatw \left(D^2 - \kxsq\right) D\right) \tvs = 0,
\end{equation}
where, $D = d/d\bary$. To obtain a stability condition, we multiply the above equation with the complex conjugate of the normal deformation, $\tvsa$, and integrate across the solid layer. The resulting equation after integration by parts becomes,
\begin{equation}
\begin{aligned}
\tvsa D^3 \tvs \Big|_{-L_r}^0 - D\tvsa D^2 \tvs \Big|_{-L_r}^0 - (2 - \lambda c^2)\kxsq \tvsa D \tvs \Big|_{-L_r}^0 + \int_{-L_r}^0 \left(|D^2 \tvs|^2 + (2 - \lambda c^2)\kxsq |D \tvs|^2 + (1 - \lambda c^2)\kx^4 |\tvs|^2 \right) d\bary\\
  - \epsilon \imath \kx c \mu_r \lambda\left( \tvsa D^3 \tvs \Big|_{-L_r}^0 - D\tvsa D^2 \tvs \Big|_{-L_r}^0 - 2 \kxsq \tvsa D \tvs \Big|_{-L_r}^0 + \int_{-L_r}^0  \left(|D^2 \tvs|^2 + 2\kxsq |D \tvs|^2 + \kx^4 |\tvs|^2 \right) d\bary  \right) \\
+ 2\epsilon \lambda \imath \kx \barsigmatw \left( \tvsa D^2 \tvs \Big|_{-L_r}^0 - \kxsq |\tvsa|^2 \Big|_{-L_r}^0 - \int_{-L_r}^0 D\tvsa\left(D^2 - \kxsq\right)\tvs d\bary  \right)= 0.
\end{aligned}
\label{eq:tvs_integral} 
\end{equation}
All the boundary terms vanish at $y = -L_r$ as $\tvs = D\tvs = 0$, and the interface conditions are used to obtain
the values of the boundary terms at $y = 0$. The procedure and corresponding algebra is detailed in appendix \ref{secapp:eig_approx}. 
After simplification, Eq. \ref{eq:tvs_integral}, at the leading order, becomes,
\begin{equation}
\begin{aligned}
\kxsq\left(\tvsazero D\tvszero + \tvszero D\tvsazero \right)\Big|_{-L_r}^0 + \int_{-L_r}^0 \left(|D^2 \tvszero|^2 + (2 - \lambda \czerosq)\kxsq |D \tvszero|^2 + (1 - \lambda \czerosq)\kx^4 |\tvszero|^2 \right) d\bary = 0.
\end{aligned}
\label{eq:tvs_integral_zero}
\end{equation} 
For $\czero = \cRzero + \imath \cIzero$, the imaginary part of the above equation can be written as,
\begin{equation}
 - 2\lambda \cRzero \cIzero \kxsq \int_{-L_r}^{0} \left(|D \tvszero|^2 + \kxsq |\tvszero|^2 \right) d\bary = 0 .
\end{equation} 
Since the integral is always positive, for modes with $\cRzero \neq 0$, $\cIzero = 0$, i.e., the modes are neutrally stable. This is expected because, at the leading order the solid momentum equations (shown in Eqs. \ref{eq:rho_mode_solid_highrhor-zero}-\ref{eq:v_mode_solid_highrhor-zero} in appendix \ref{secapp:asymp_eqs}) are devoid of the viscous dissipation part and the elastic stresses are always conservative. It is easy to see from Eq. \ref{eq:vs_mode_1} that the leading order evolution equation is free of any imaginary component, and so do the interface conditions shown in Eq. \ref{eq:d3tvszero} and \ref{eq:d2tvszero}. Therefore, we choose to scale the leading order eigenfunctions such that $\tvs$ is real.   

As the leading order solution is real, the stability of the modes can be determined at the next order. In order to obtain $\cone$, we write the $O(\epsilon)$ term of Eq. \ref{eq:tvs_integral}. Since the leading order instability will be determined only by the imaginary part of $\cone$, we take the imaginary part of that equation. After substitution of the boundary terms and simplification (algebra detailed in appendix \ref{secapp:eig_approx}), the resulting equation obtained can be written as, 
\begin{equation}
\begin{aligned}
- \lambda\kxsq \tvsazero &\mbox{Imag}(\tsigmayyzero)\Big|_{y=0} + \kx\lambda D\tvsazero \mbox{Real}\left(\tsigmaxyzero + D(\barmu D\baru) \tvszero \right)\Big|_{y=0} - \lambda\kx\barsigmatw \left( |D\tvszero|^2 - \kxsq|\tvszero|^2 \right)\Big|_{y=0}\\
- & 2\lambda \czero \cIone \kxsq \int_{-L_r}^0 \left(|D \tvszero|^2 + \kxsq |\tvszero|^2 \right) d\bary  \\
& - \kx\czero\lambda\mu_r \left( 2\kxsq \tvsazero D\tvszero\Big|_{y=0} + \int_{-L_r}^0  \left(|D^2 \tvszero|^2 + 2\kxsq |D \tvszero|^2 + \kx^4 |\tvszero|^2 \right) d\bary \right) = 0,
\end{aligned}
\label{eq:tvs_integral_zero_simpl}
\end{equation}
where, 
\begin{equation}
\begin{aligned}
 \tsigmayyzero &= -\frac{\Re\tpzero}{\gamma\Ma^2} + {2\barmu}\frac{d\tvzero}{dy}  - \frac{2\barmu}{3}\left(\imath\kx\tuzero + \frac{d\tvzero}{dy} \right), \\
 \tsigmaxyzero &= {\barmu}\left(\frac{d\tuzero}{dy} + \imath\kx\tvzero \right) + \frac{d\mu}{dT}\Bigg|_{\barT} \frac{d\baru}{dy} \tTzero.
\end{aligned}
\label{eq:mode_str_zero}
\end{equation}
Rearranging the terms, we obtain the expression for the leading order imaginary part of the wave-speed,
\begin{equation}
\begin{aligned}
 I_0 \cIone = -\kx \tvsazero\mbox{Imag}(\tsigmayyzero)\Big|_{y=0} &+ D\tvsazero \mbox{Real}\left(\tsigmaxyzero + D(\barmu D(\baru)) \tvszero \right)\Big|_{y=0} - \barsigmatw \left( |D\tvszero|^2 - \kxsq|\tvszero|^2 \right)\Big|_{y=0} \\
 & - \mu_r \czero \left( 2\kx^2 \tvsazero D\tvszero\Big|_{y=0} + \int_{-L_r}^0  \left(|D^2 \tvszero|^2 + 2\kxsq |D \tvszero|^2 + \kx^4 |\tvszero|^2 \right) d\bary \right) ,
\end{aligned}
\label{eq:coneI}
\end{equation}
where,
\begin{equation}
 I_0 = {2\kx\czero \int_{-L_r}^0 \left(|D \tvszero|^2 + \kxsq |\tvszero|^2 \right) d\bary}.
\label{eq:def_I0}
\end{equation}
Therefore, the leading order imaginary part of the eigenvalue corresponding to the elastic modes can be expressed as a sum of the terms on the right hand side of Eq. \ref{eq:coneI}. The first two terms represent the contribution of the fluid stresses at the interface, the third term represents the interfacial contribution of the non-linearity in the hyperelastic solid model, and the fourth term is the contribution due to the dissipation in the solid, which comprises of a boundary contribution and an integral across the solid domain.

\begin{figure}
\begin{center}
\begin{minipage}{.45\textwidth}
  \includegraphics[width=3in]{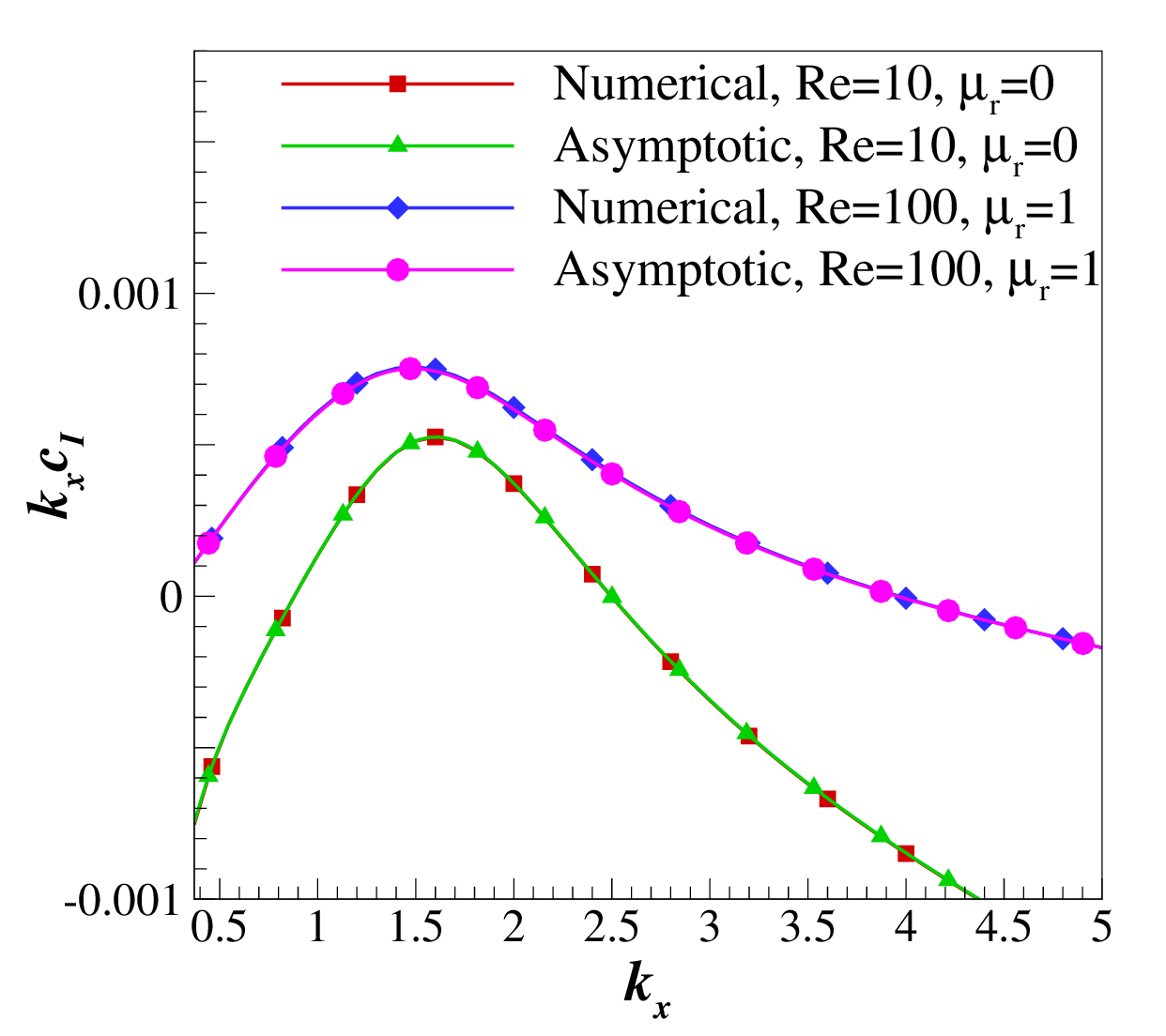}
  \caption{Comparison of growth rates obtained through numerical computation and asymptotic solution ($\cI = (\rho_r\Re)^{-1}\cIone$) of the first elastic mode for two cases: Case I - $\Re = 10$, $\mu_r = 0$, $\lambda = 100$, Case II - $\Re = 100$, $\mu_r = 1$, $\lambda = 1000$. For both cases, $\rho_r = 1000$, $\Ma = 1$, bottom wall is adiabatic with $\kappa_r = 100$.}
  \label{fig-disp_asymp_re}
\end{minipage}
\quad
\begin{minipage}{.45\textwidth}
  \includegraphics[width=3in]{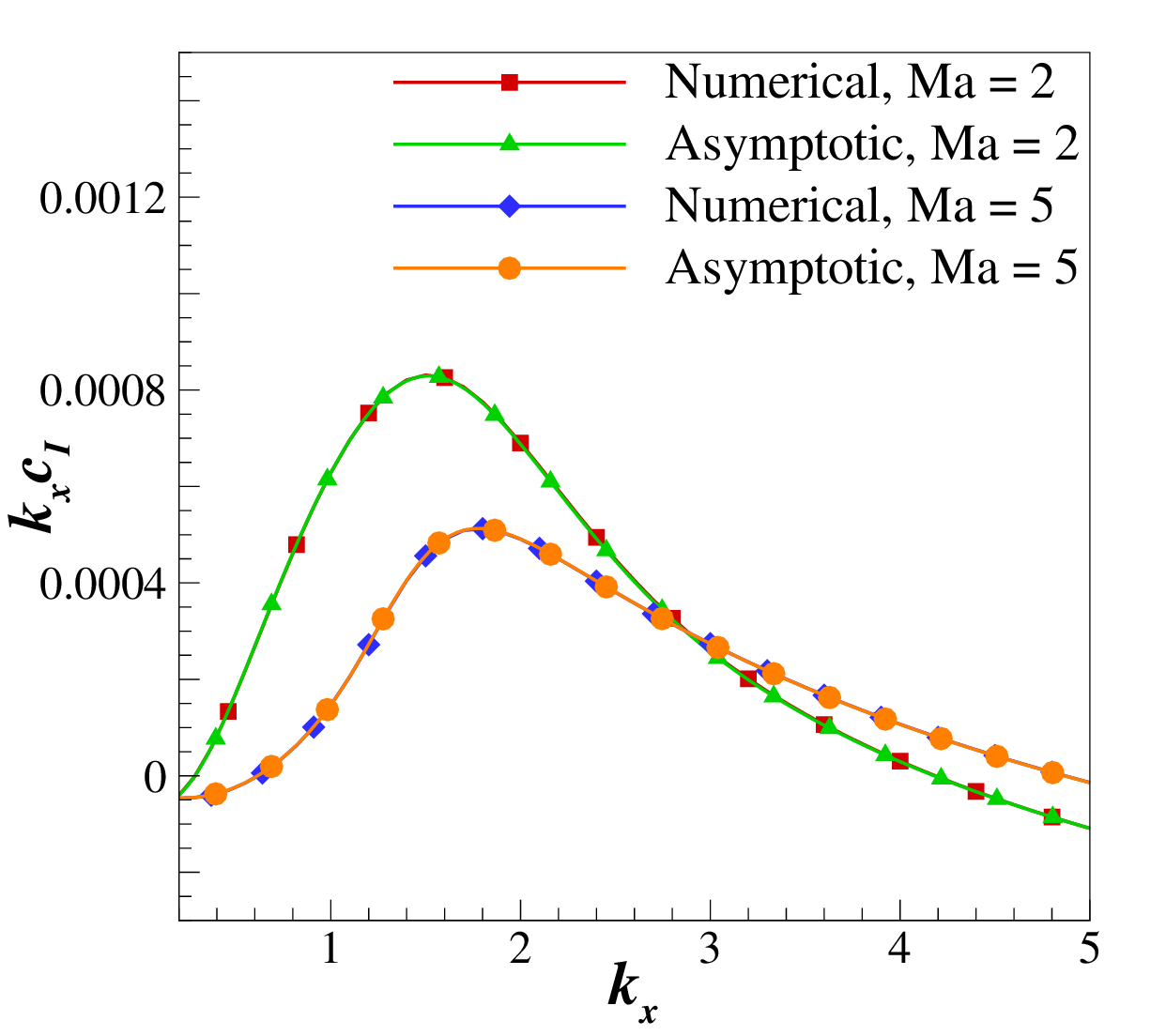}
  \caption{Comparison of growth rates obtained through numerical computation and asymptotic solution ($\cI = (\rho_r\Re)^{-1}\cIone$) of the first elastic mode for two cases: Case I - $\Ma = 2$ and bottom wall is adiabatic, Case II - $\Ma = 5$ and bottom wall is isothermal with $T_{bw} = 1$. For both cases, $\rho_r = 1000$, $\Re = 100$, $\mu_r = 0$, and, $\kappa_r = 100$.}
  \label{fig-disp_asymp}
\end{minipage}
\end{center}
\end{figure}

To test the accuracy of the asymptotic result, we compute the growth rates numerically and compare against the results from the asymptotic expression, Eq. \ref{eq:coneI} (as $\cIzero = 0$, therefore, $\cI = \epsilon \cIone$). Figure \ref{fig-disp_asymp_re} shows the comparison at $\Ma = 1$, for two different values of $\Re$, $\lambda$, and $\mu_r$, while figure \ref{fig-disp_asymp} shows the comparison for $\Ma = 2$ and $5$, for an adiabatic bottom wall and an isothermal bottom wall case. In all cases, the asymptotic solutions show an excellent match with the numerical solutions, thereby, demonstrating the efficacy of the asymptotic analysis. 

\subsection{Interpretation of the interface contribution terms from the flow}

The leading order imaginary part of the instability is dictated by the values of $\tsigmaxyzero$ and $\tsigmayyzero$ at the interface, which are effectively the tangential and normal stresses exerted by the flow at the solid-fluid interface. It is possible to express these quantities as integrals of the solution in the flow. To do so, we re-write the momentum equations in the flow, Eqs. \ref{u_mode}-\ref{v_mode} in appendix \ref{secapp:modal_eqs}, as,
\begin{equation}
 \barrho \left(\imath \kx ( \baru - \czero ) \tuzero + \frac{d\baru}{dy} \tvzero \right) = \frac{\imath\kx\tsigmaxxzero}{\Re} + \frac{1}{\Re}\frac{d\tsigmaxyzero}{dy},
\label{eq:u_mode_alt}
\end{equation}  
\begin{equation}
 \barrho \left(\imath \kx ( \baru - \czero ) \tvzero \right) = \frac{\imath\kx \tsigmaxyzero}{\Re} + \frac{1}{\Re}\frac{d\tsigmayyzero}{dy},
\label{eq:v_mode_alt}
\end{equation}  
where, 
\begin{equation}
 \tsigmaxxzero = -\frac{\Re\tpzero}{\gamma\Ma^2} + {2\barmu }\imath \kx\tuzero - \frac{2\barmu}{3}\left(\imath\kx\tuzero + \frac{d\tvzero}{dy} \right),
\label{eq:mode_strxx_zero}
\end{equation}
and, $\tsigmaxyzero$ and $\tsigmayyzero$ as defined in Eq. \ref{eq:mode_str_zero}. First, we multiply Eq. \ref{eq:u_mode_alt} with $\tuazero$ and Eq. \ref{eq:v_mode_alt} with $\tvazero$ and add the two equations; then integrate it across the fluid domain. Keeping in mind that $\czero$ is real for the elastic modes, the imaginary part of the resulting equation simplifies to,
\begin{equation}
\begin{aligned}
 \int_0^1 \barrho \frac{d\baru}{dy} \mbox{Real}\left(\tuazero \tvzero\right) dy = \frac{1}{\Re}\mbox{Real}&\left(-\left(\tuazero\tsigmaxyzero + \tvazero\tsigmayyzero \right)\Big|_{y=0}\right) \\
 &+ \frac{1}{\Re}\int_0^1\mbox{Real}\left(\imath\kx\tsigmaxxzero - \tsigmaxyzero\frac{d\tuazero}{dy} + \imath\kx\tsigmaxyzero\tvazero - \tsigmayyzero\frac{d\tvzero}{dy}\right)dy.
\end{aligned}
\label{eq:budget_imag}
\end{equation}
The second term on the right hand side of the above expression is an integral across the fluid domain and can be evaluated by substituting $\tsigmaxxzero$, $\tsigmaxyzero$ and $\tsigmayyzero$, from Eqs. \ref{eq:mode_strxx_zero} and \ref{eq:mode_str_zero}, respectively. Upon substitution and rearrangement of the terms, we get, 
\begin{equation}
\begin{aligned}
\frac{1}{\Re}\left(\imath\kx\tsigmaxxzero - \tsigmaxyzero\frac{d\tuazero}{dy} + \imath\kx\tsigmaxyzero\tvazero - \tsigmayyzero\frac{d\tvzero}{dy}\right) = \frac{\tpzero \divazero}{\gamma\Ma^2} -\frac{1}{\Re}\frac{d\mu}{dT}\Big|_{\barT}\frac{d\baru}{dy} \tTzero \ttauxyazero - \tPhizero,
\end{aligned}
\label{eq:energy_balance_terms}
\end{equation}
where,
\begin{equation}
\begin{aligned}
\divzero &= \imath\kx\tuzero + \frac{d\tvzero}{dy} ,\:\:\:\ttauxyzero =  \frac{d\tuzero}{dy} + \imath\kx\tvzero , \\
\tPhizero &= \frac{\barmu}{\Re}\left(\left|\frac{d\tuzero}{dy} + \imath\kx\tvzero\right|^2 + 2\left(\left|\frac{d\tvzero}{dy}\right|^2 + \kxsq\left|\tuzero\right|^2 \right) - \frac{2}{3}\left|\imath\kx\tuzero + \frac{d\tvzero}{dy}\right|^2\right).
\end{aligned}
\label{eq:energy_balance_terms_expansion}
\end{equation}
The first term on the right hand side of Eq. \ref{eq:budget_imag} is evaluated at the interface. Using the velocity continuity conditions (Eqs. \ref{eq:ic1_mode_incomp_highrhor-zero}-\ref{eq:ic2_mode_incomp_highrhor-zero}, appendix \ref{secapp:asymp_eqs}) it is easy to show that,
\begin{equation}
\mbox{Real}\left(\tuazero\tsigmaxyzero + \tvazero\tsigmayyzero\right)\Big|_{y=0} = \czero\left( -\kx \tvsazero \mbox{Imag}\left(\tsigmayyzero\right) + D\tvsazero \mbox{Real}\left(\tsigmaxyzero\right)\right)\Big|_{y=0} - D\baru\tvsazero\mbox{Real}(\tsigmaxyzero)\Big|_{y=0}.
\label{eq:bcontr}
\end{equation}
where, the quantity inside the parentheses of the first term on the right hand side appears as the interfacial fluid stress contribution in the expression for $\cIone$ (Eq. \ref{eq:coneI}). Therefore, substituting Eqs. \ref{eq:bcontr} and \ref{eq:energy_balance_terms} into Eq. \ref{eq:budget_imag} and rearranging, the interfacial fluid stress contribution term can be written as,
\begin{equation}
\begin{aligned}
\Big( -\kx \tvsazero \mbox{Imag}\left(\tsigmayyzero\right) + &D\tvsazero \mbox{Real}\left(\tsigmaxyzero\right)\Big)\Big|_{y=0} = \int_0^1 -\frac{\barrho}{\czero} \frac{d\baru}{dy} \mbox{Real}\left(\tuazero \tvzero\right) dy + \int_0^1 \frac{\mbox{Real}\left(\tpzero\divazero\right)}{\czero\gamma\Ma^2}dy \\
& + \int_0^1 -\frac{1}{\czero\Re}\frac{d\mu}{dT}\Big|_{\barT}\frac{d\baru}{dy} \mbox{Real}\left(\tTzero \ttauxyazero\right) dy - \int_0^1 \frac{\tPhizero}{\czero} dy + \frac{D\baru}{\czero}\tvsazero\mbox{Real}(\tsigmaxyzero)\Big|_{y=0}.
\end{aligned}
\end{equation}
Substituting the above expression into the flow contribution terms of Eq. \ref{eq:coneI}, we can re-write the expression for $\cIone$ in a compact form as,
\begin{equation}
 \cIone = I_1 + I_2 + I_3 + I_4 + B_1 + B_2 + B_3 - \mu_r(B_4 + I_5),
\label{eq:ci_compact}
\end{equation}
where, 
\begin{equation}
\begin{aligned}
 I_1 &= -\frac{\Re}{I_0 \czero}\int_0^1 \barrho \frac{d\baru}{dy} \mbox{Real}\left(\tuazero \tvzero\right) dy,\:\:\:
 I_2 =  \frac{\Re}{I_0 \czero}\int_0^1 \frac{\mbox{Real}\left(\tpzero\divazero\right)}{\gamma\Ma^2}dy,\\
 I_3 &=  -\frac{1}{I_0 \czero} \int_0^1 \frac{d\mu}{dT}\Big|_{\barT}\frac{d\baru}{dy} \mbox{Real}\left(\tTzero \ttauxyazero\right) dy,\:\:\:
 I_4 =  -\frac{\Re}{I_0 \czero} \int_0^1 \tPhizero dy,\\
 I_5 &= \frac{\czero}{I_0} \int_{-L_r}^0  \left(|D^2 \tvszero|^2 + 2\kxsq |D \tvszero|^2 + \kx^4 |\tvszero|^2 \right) d\bary,\\
 B_1 &= \frac{D\baru}{I_0 \czero}\tvsazero\mbox{Real}(\tsigmaxyzero)\Big|_{y=0},\:\:\:
 B_2 = \frac{D(\barmu D\baru)}{I_0} D\tvsazero\tvszero \Big|_{y=0},\:\:\:
 B_3 = - \frac{\barmu D\baru}{I_0} \left( |D\tvszero|^2 - \kxsq|\tvszero|^2 \right)\Big|_{y=0},\\
 B_4 &=  \frac{2\kx^2\czero}{I_0} \tvsazero D\tvszero\Big|_{y=0},
\end{aligned}
\label{eq:ci_compact_terms}
\end{equation}
and $I_0$ is shown in Eq. \ref{eq:def_I0}. Each term contributing to the leading order instability carries a physical meaning. The first four terms, $I_1-I_4$ are integrals in the fluid domain that represent the shear production, pressure dilatation, action of shear strain on viscosity perturbation, and viscous dissipation, respectively. For an incompressible constant-viscosity flow, $I_2$ and $I_3$ are identically zero, but they are important for a compressible flow. The term $I_5$ is an integral across the solid domain that is active for dissipative solids. The first two boundary terms, $B_1$ and $B_2$ signify the effect of interfacial deformation that manifests through the action of tangential velocity on shear stress, whereas $B_3$ is an interfacial effect of the non-linearity in the constitutive model of the solid. $B_4$ is a boundary term that appears due to the dissipative term in the solid. It is prudent to note that the effect of the fluid perturbations on the leading order instability is felt through the terms, $I_1-I_4$ and $B_1$, while the remaining terms are purely functions of the perturbations fields in the solid.

\begin{figure}
\begin{center}
\begin{minipage}{.45\textwidth}
  \includegraphics[width=3in]{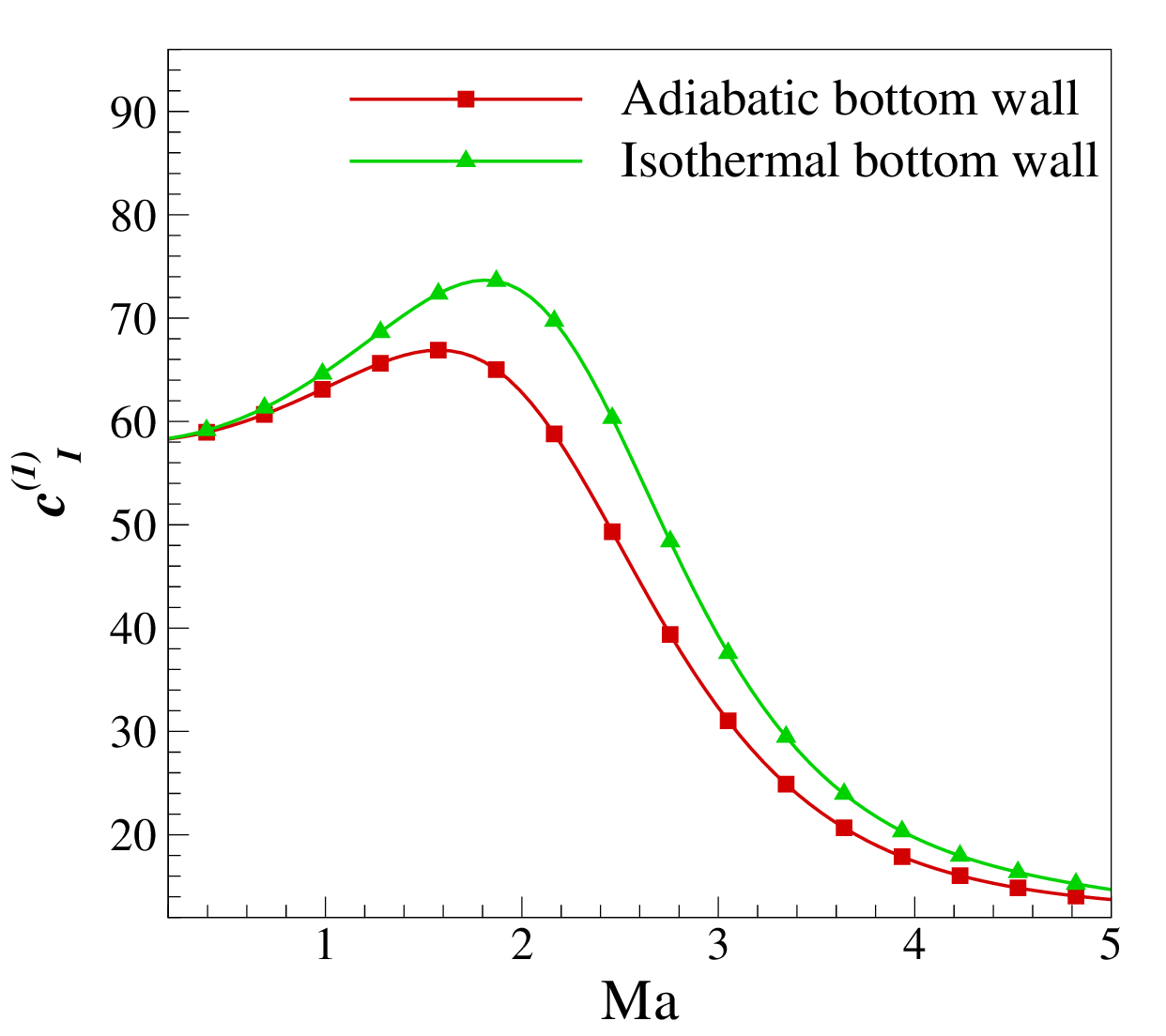}
  \caption{Leading order imaginary part of wave-speed ($\cIone$) as a function of Mach number for adiabatic bottom wall and isothermal bottom wall ($T_{bw} = 1$) cases, computed at $\kx=1$, $\Re = 100$, $\lambda = 1000$, $\rho_r = 1000$, $\mu_r = 0$, $\kappa_r = 100$. }
  \label{fig-ci-ma}
\end{minipage}
\quad
\begin{minipage}{.45\textwidth}
  \includegraphics[width=3in]{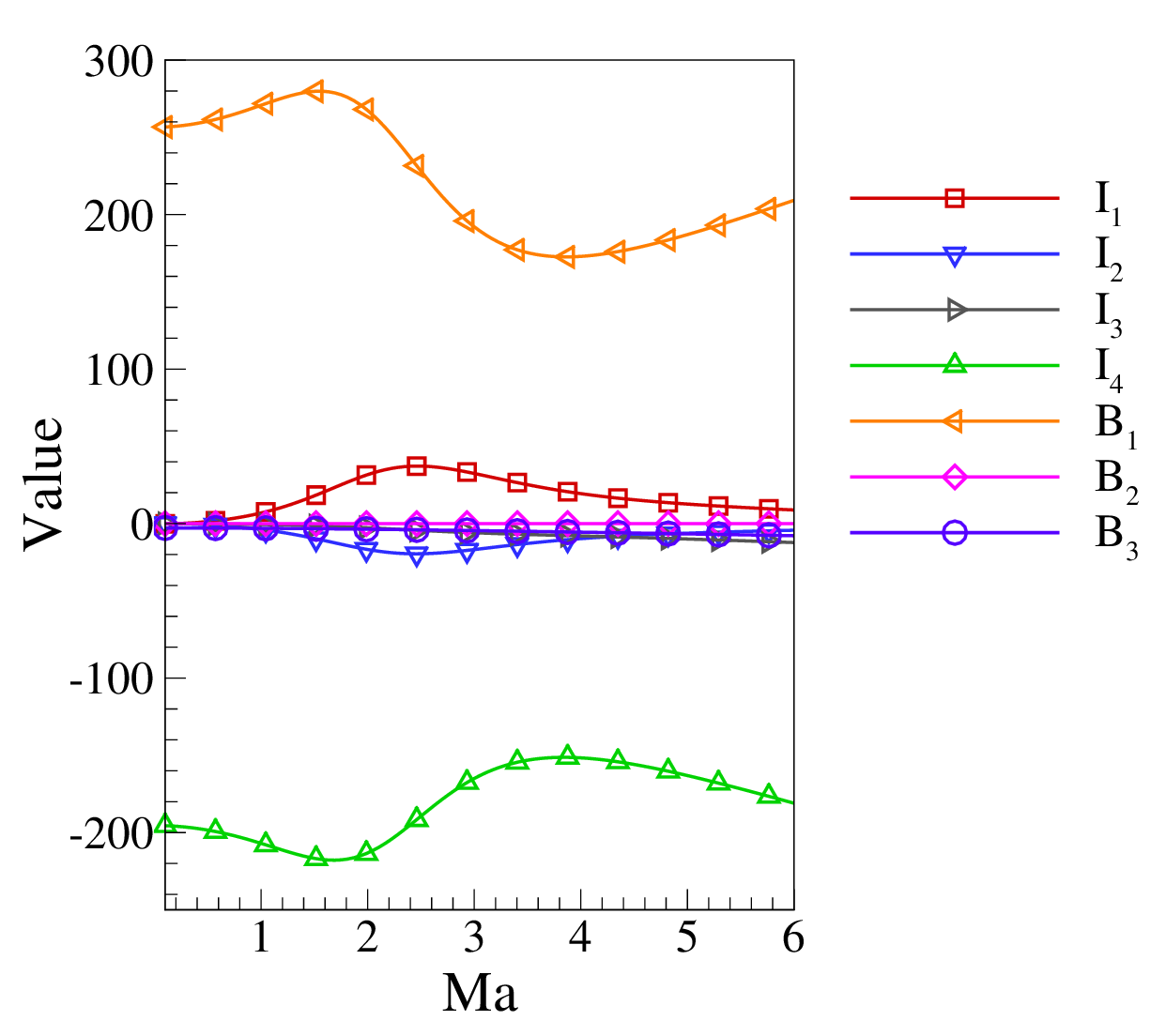}
  \caption{Variation of the terms on the right hand side of Eq. \ref{eq:ci_compact} with Mach number for the adiabatic bottom wall case computed at $\kx=1$, $\Re = 100$, $\lambda = 1000$, $\rho_r = 1000$, $\mu_r = 0$, $\kappa_r = 100$.}
  \label{fig-budget-ma}
\end{minipage}
\end{center}
\end{figure}

\subsection{Effect of different parameters on the leading order instability}

\begin{figure}
\begin{center}
\begin{minipage}{.45\textwidth}
  \includegraphics[width=3in]{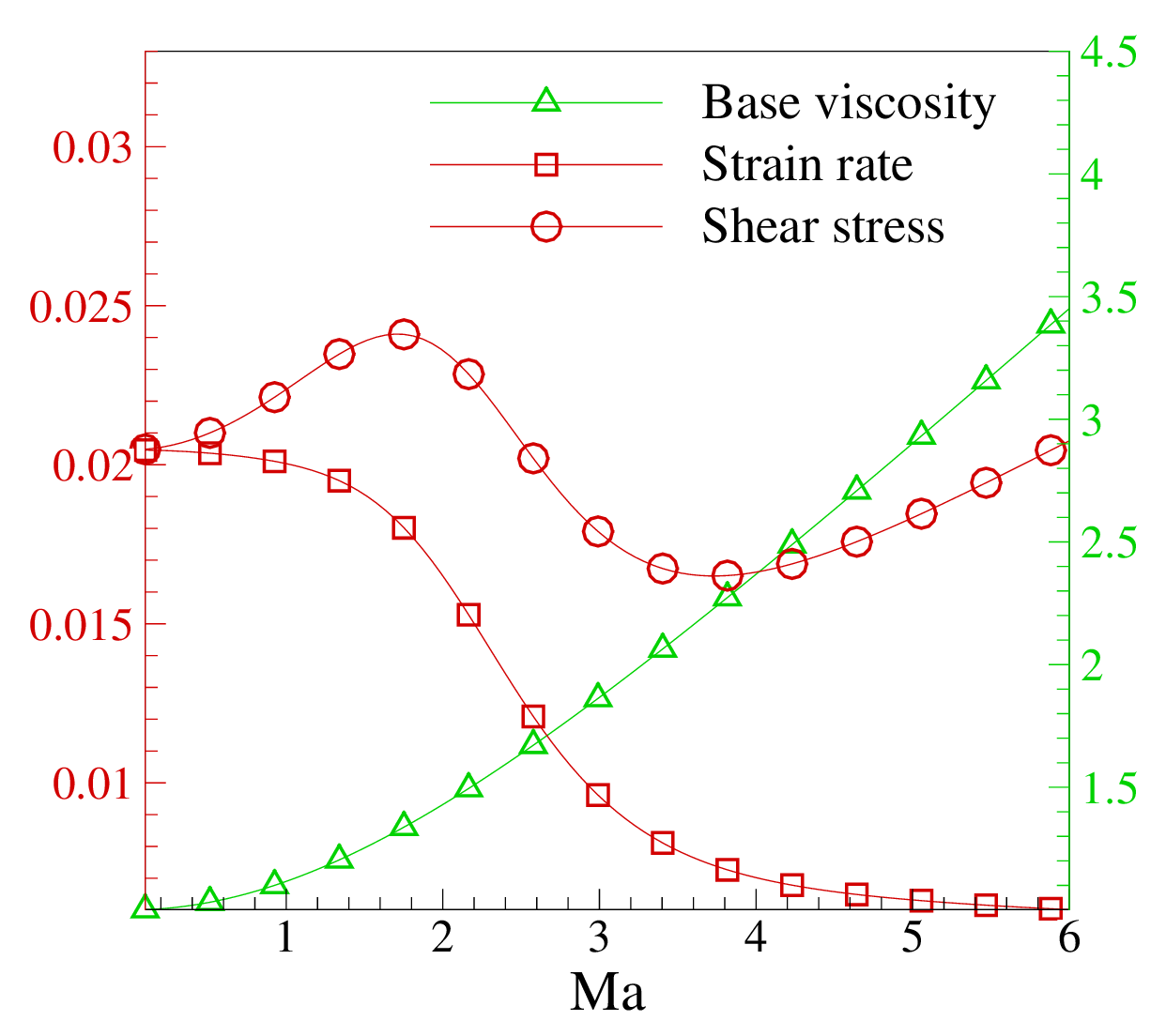}
  \caption{Variation of the base flow viscosity ($\barmu$), perturbation shear strain ($\ttauxyzero = d\tuzero/dy + \imath\kx\tvzero$) and perturbation shear stress ($\tsigmaxyzero = \barmu \ttauxyzero$), at $y=0$, with Mach number for the adiabatic bottom wall case, computed at $\kx=1$, $\Re = 100$, $\lambda = 1000$, $\rho_r = 1000$, $\mu_r = 0$, $\kappa_r = 100$. }
  \label{fig-wall-str}
\end{minipage}
\quad
\begin{minipage}{.45\textwidth}
  \includegraphics[width=3in]{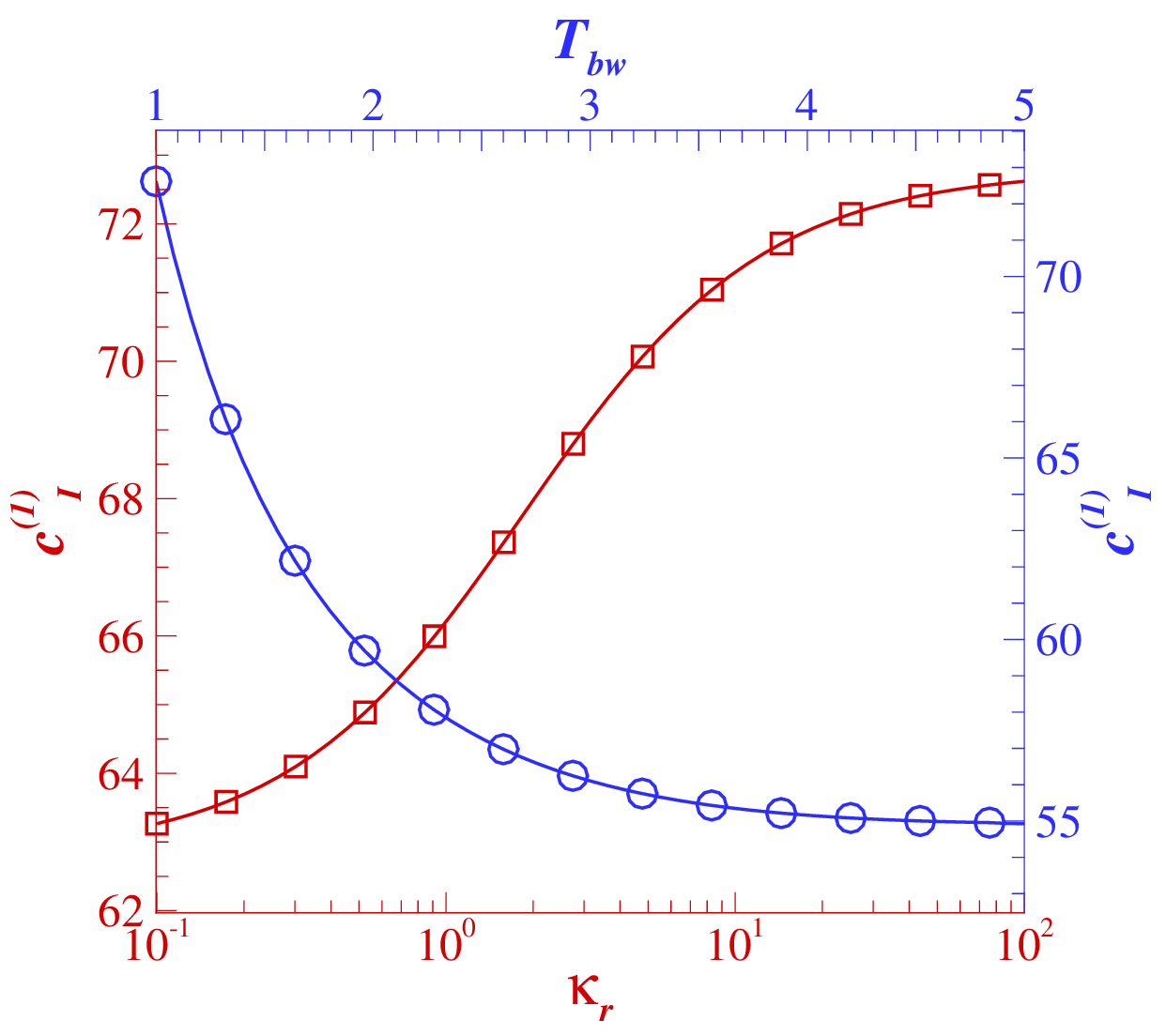}
  \caption{Leading order imaginary part of wave-speed ($\cIone$) as a function of conductivity ratio ($\kappa_r$), shown with red square symbol ($\textcolor{red}{\square}$), and bottom wall temperature ($T_{bw}$), shown with the blue circle symbol ($\textcolor{blue}{\bigcirc}$), computed at $\kx=1$, $\Ma = 2$, $\Re = 100$, $\lambda = 1000$, $\rho_r = 1000$, $\mu_r = 0$, $\kappa_r = 100$ (for the $T_{bw}$ variation plot) and $T_{bw} = 1$ (for the $\kappa_r$ variation plot). }
  \label{fig-kappar-tbw-var}
\end{minipage}
\end{center}
\end{figure}

In the previous section, we obtained a closed-form expression for the leading order imaginary part of the eigenvalue that dictates the stability of a flow past a compliant solid. We can, therefore, compute each of the contributions to $\cIone$ and plot it for different values of relevant non-dimensional parameters to study the effect of each term on the instability. The first parameter we study is the Mach number, which characterizes the degree of compressibility of the flow. Figure \ref{fig-ci-ma} shows the variation of $\cIone$, computed using Eqs. \ref{eq:ci_compact} and \ref{eq:ci_compact_terms}, with $\Ma$ for the adiabatic and isothermal bottom wall cases. Both cases show a non-monotonic behaviour, with $\cIone$ increasing initially with increase in Mach number followed by a decrease. This is consistent with the observations made from the numerical calculations (figure \ref{fig-disp_rel_ma} in section \ref{sec:results}). Additionally, $\cIone$ is higher for the isothermal bottom wall case compared to the adiabatic counterpart, which is also consistent with the numerical observations made through figure \ref{fig-disp_rel_kappar} in section \ref{sec:results}. To understand the non-monotonic variation with Mach number, we plot the variation of the different terms that add up to $\cIone$ (from Eq. \ref{eq:ci_compact}), shown in figure \ref{fig-budget-ma}. From the figure it is evident, that the primary contribution to the leading order instability is due to the terms, $B_1$, $I_4$ and $I_1$. While, $I_4$, which denotes the viscous dissipation has a clear stabilizing influence at all Mach numbers, the interfacial stress work term $B_1$, and the shear production term $I_1$ are observed to be always positive and hence destabilizing in nature. It is also seen that $B_1$ peaks at about $\Ma = 2$ and then starts to decrease before saturating, whereas the opposite behaviour is seen for $I_2$. At the same time, the shear production term increases to a peak at about Mach number of about $2-3$. The combined effect of the three terms leads to a peak in the value of $\cIone$ around Mach number of $2$. To further examine this, if we note the interfacial stress work term $B_1$, shown in Eq. \ref{eq:ci_compact_terms}, comprises of the real part of the leading order fluid shear stress ($\tsigmaxyzero$) at the interface, which can be written as, $\tsigmaxyzero = \barmu(d\tuzero/dy + \imath\kx\tvzero)$, which is the product of the base viscosity, $\barmu$, and the shear strain, $\ttauxyzero$, defined in Eq. \ref{eq:energy_balance_terms_expansion}. Figure \ref{fig-wall-str} shows the variation of $\tsigmaxyzero, \ttauxyzero$ and, $\barmu$, at the interface with increasing Mach number. From the figure it is evident that with increasing Mach number, the base viscosity shows a monotonic increase whereas the base strain rate shows a decreasing trend. The result is that the product of the two, which is the shear stress, shows a non-monotonic variation with a peak occurring at about $\Ma = 2$, which is roughly the same trend observed in $\cIone$ vs. $\Ma$ curve.

\begin{figure}
\begin{center}
\begin{minipage}{.45\textwidth}
  \includegraphics[width=3in]{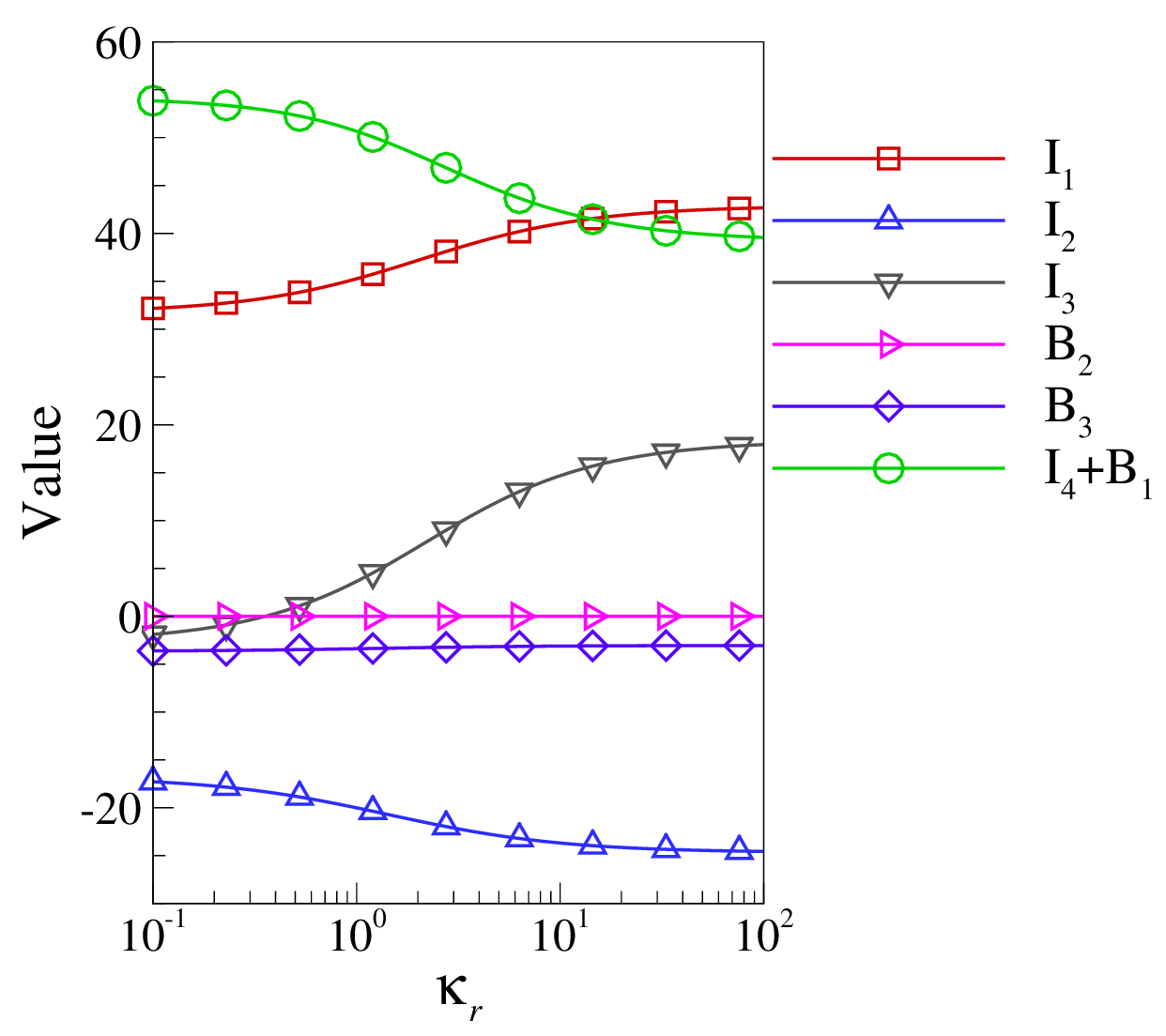}
  \caption{Variation of the terms on the right hand side of Eq. \ref{eq:ci_compact} with conductivity ratio ($\kappa_r$) for the isothermal bottom wall case computed at $\kx=1$, $\Ma = 2$, $\Re = 100$, $\lambda = 1000$, $\rho_r = 1000$, $\mu_r = 0$, $T_{bw} = 1$.}
  \label{fig-budget-kappar}
\end{minipage}
\quad
\begin{minipage}{.45\textwidth}
  \includegraphics[width=3in]{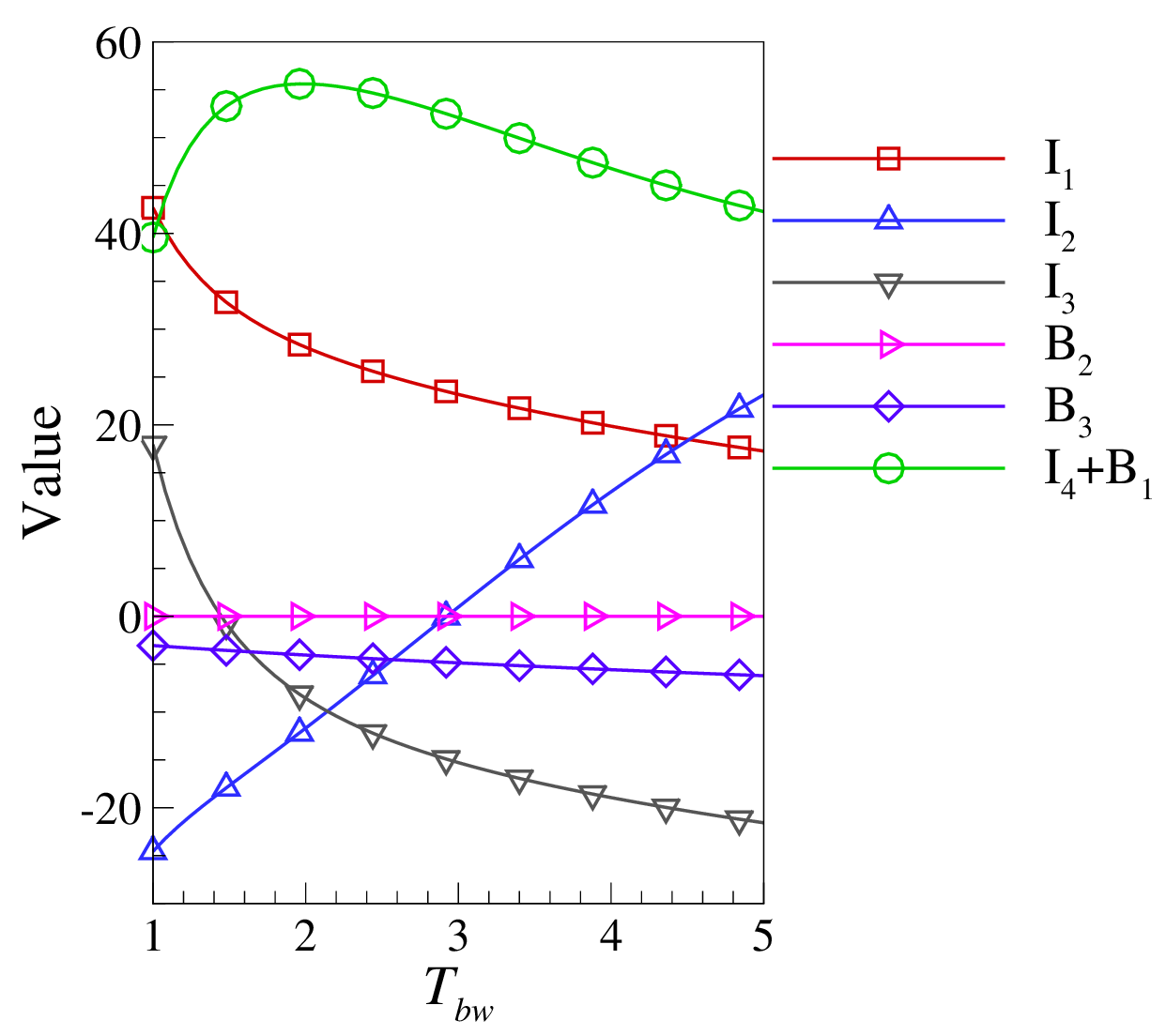}
  \caption{Variation of the terms on the right hand side of Eq. \ref{eq:ci_compact} with bottom wall temperature ($T_{bw}$) for the isothermal bottom wall case computed at $\kx=1$, $\Ma = 2$, $\Re = 100$, $\lambda = 1000$, $\rho_r = 1000$, $\mu_r = 0$, $\kappa_r = 100$.}
  \label{fig-budget-tbw}
\end{minipage}
\end{center}
\end{figure}

Besides the flow compressibility, two other factors that influence the stability of the elastic modes for compressible flows are the conductivity ratio ($\kappa_r$) and the bottom-wall temperature ($T_{bw}$). From figure \ref{fig-disp_rel_kappar} in section \ref{sec:results}, it is evident that the $\kappa_r$ has negligible effect on the stability of the elastic modes for an adiabatic bottom wall case, which as justified in section \ref{sec:results}, is due to the base flow profile being unchanged. For an isothermal bottom wall case, that is not the case. Figure \ref{fig-kappar-tbw-var} show $\cIone$ vs. $\kappa_r$ for the isothermal bottom wall case with $T_{bw} = 1$ using the square ($\square$) symbol, and $\cIone$ vs. $T_{bw}$ for $\kappa_r=100$ using the circle ($\bigcirc$) symbol. From the figure it is evident that increasing $\kappa_r$ destabilizes the system whereas increasing bottom wall temperature has a stabilizing effect. Both observations are, again, consistent with the numerical results shown in figures \ref{fig-disp_rel_kappar} and \ref{fig-disp_rel_Tbw} in section \ref{sec:results}. To understand the influence of the different factors, we plot the terms constituting the right hand side of Eq. \ref{eq:ci_compact} as a function of $\kappa_r$ and $T_{bw}$. Similar to the variation with Mach number, in both cases, the most dominant destabilizing and stabilizing influences are interfacial stress work term, $B_1$, and the viscous dissipation term, $I_4$, respectively (not shown here for the sake of brevity). However, the sum of their effects are comparable to the other terms, which is plotted alongside the other terms in figures \ref{fig-budget-kappar} and \ref{fig-budget-tbw}. From figure \ref{fig-budget-kappar}, it can be seen that the sum of the two dominant terms, $I_4 + B_1$ decreases with increasing $\kappa_r$, and so does the pressure dilatation term $I_2$. However, there is a monotonic increase of the shear production term $I_1$ and the term $I_3$, which denotes the action of shear strain on viscosity perturbations. The overall effect is a monotonic increase in $\cIone$ with increasing $\kappa_r$. Similarly, in figure \ref{fig-budget-tbw}, we plot $I_4 + B_1$ and the remaining terms as a function of $T_{bw}$. Here the only term that monotonically increases with increasing bottom wall temperature is the pressure dilatation term, $I_2$. Besides that all terms have a decreasing trend which overall leads to stabilization of the elastic modes. The trends of the term $I_3$ can be explained by the fact viscosity perturbations are directly proportional to the temperature gradients. As evident from the base profiles shown in figure \ref{fig-base_kappar_var}(b), the temperature gradients increase with increasing conductivity ratio and would decrease with increasing bottom wall temperature. This causes the term $I_3$ to monotonically increase with increasing $\kappa_r$ and decrease with increasing $T_{bw}$. In this way, the effect of the different mechanisms on the instability can be isolated as a function of the relevant non-dimensional parameters.

\section{Conclusion and Discussion}
\label{sec:conclusions}

In this work, we study the linear-modal-temporal stability of a gas flow past a viscoelastic compliant layer anchored to a rigid base. The configuration is modeled as a shear-driven, compressible flow of a calorifically perfect gas over a Neo-Hookean solid with viscous damping. A defining characteristic of this system is the high solid-to-fluid density ratio ($\rho_r$). Numerical calculations reveal that at high $\rho_r$, the dominant instability are the elastic modes of the solid. These elastic modes manifest as shear waves propagating through the solid medium, which are subsequently destabilized by fluid stresses acting across the deformable solid-fluid interface.


For the regime where $\rho_r\Re \gg 1$ -- highly characteristic of applications in aerospace and biological systems -- we performed an asymptotic analysis using $\epsilon = (\rho_r\Re)^{-1}$ as a small parameter. This analysis demonstrates that the fluid-solid system can be sequentially decoupled at each asymptotic order, yielding several key implications:
\begin{itemize}
\item The leading-order elastic modes emerge as solutions to a linear elastic model subject to a free-shear condition at the interface. Because the additional stress terms characteristic of the non-linearity in the Neo-Hookean model scale as $O(\epsilon)$ -- owing to their dependence on the base-state shear, which is inherently small at high $\rho_r$ -- extending this asymptotic framework to alternative hyperelastic formulations (such as Mooney-Rivlin or Ogden models) is mathematically straightforward.
\item System instability is dictated by solving the $O(\epsilon)$ correction to the solid equations, which requires evaluating the leading-order fluid stresses at the compliant interface. We derived a closed-form expression for the leading-order imaginary part of the eigenvalues, directly isolating the contributions of these interfacial fluid stresses. Consequently, by supplying the appropriate interfacial stresses, this analysis can be readily adapted to both incompressible or other compressible plane-parallel flows.
\item In the closed-form expression for the leading-order imaginary part of the eigenvalue, the interfacial fluid stress contributions can be recast as integrals of the flow solution that carry distinct physical meaning. This framework provides a rigorous, quantitative tool to isolate and interpret the underlying physical mechanisms driving the destabilization of elastic modes in gas-solid systems.
\end{itemize}

Additionally, this study maps the effect of flow compressibility on the stability of the elastic modes, revealing a distinct, non-monotonic impact. Increasing the Mach number destabilizes the system up to $\Ma\approx 2$, beyond which further increase in $\Ma$ promotes stabilization. The asymptotic analysis attributes this behavior to the competition between interfacial shear work (driven by surface deformation) and viscous dissipation. Specifically, the interfacial shear work depends on the interfacial shear stress, which exhibits a non-monotonic trend driven by the product of a monotonically increasing base-flow viscosity and a decreasing interfacial shear strain.

Finally, because compressible flows inherently couple the momentum and energy equations in the fluid -- and subsequently bridge the temperature fields across the solid and fluid domains -- the thermal characteristics of the solid play a pivotal role in system stability. For a configuration where the bottom wall, where the compliant layer is pinned, is maintained at a constant temperature (isothermal bottom wall), the base-flow temperature profile transitions from an adiabatic-like to an isothermal-like profile as the solid-to-fluid thermal conductivity ratio ($\kappa_r$) spans from very small to very large values. As a consequence, the temporal growth rates of the elastic modes amplify with increasing $\kappa_r$. Our asymptotic framework tracks this destabilization primarily to enhanced shear production and the action of shear strain on viscosity perturbations. Conversely, elevating the bottom wall temperature altogether stabilizes the elastic modes.

Beyond its theoretical value, the framework established in this work serves as a vital predictive tool for a broad spectrum of engineering and physiological systems where the solid-to-fluid density gap is large. The findings of this work offer a foundational step toward understanding and modeling laminar-to-turbulent transition mechanisms in high-speed gas flows over compliant substrates. In aerospace applications -- where structures routinely experience extreme aerothermal loads through high-speed boundary layer flows -- the ability to accurately predict fluid-structure instabilities is paramount. Because the closed-form asymptotic solutions derived herein hold exceptionally well in these high-density-ratio regimes ($\rho_r\Re \gg 1$), they bypass the need for computationally prohibitive, fully coupled numerical simulations during early-stage design phases. Furthermore, the inherent modularity of this framework allows it to be readily extended to alternative hyperelastic structural models and alternate parallel flow configurations. On a physiological scale, this analysis is equally applicable to biological transport phenomena, such as human respiratory mechanics. In these systems, the interaction between airflow and highly deformable, viscoelastic mucosal or tissue layers dictates airway patency and stability, potentially making the physics-based insights of this model relevant for biomedical engineering diagnostics.

\section*{Acknowledgements}   
The author gratefully acknowledges the Department of Science and Technology (DST), Ministry of Science and Technology, Government of India, for financial support through the DST INSPIRE Faculty Fellowship (IFA24-ENG 379). The author also extends sincere thanks to Prof. V. Kumaran for his insightful discussions and support during the initial phase of this work.

\appendix
\section{Normal mode equations}
\label{secapp:modal_eqs}

The governing equations for the two-dimensional normal modes in the fluid, are,
\begin{equation}
\label{rho_mode}
 \imath \kx ( \baru - c ) \trho + \frac{d \barrho}{dy} \tv + \barrho \left( \imath \kx \tu + \frac{d \tv}{dy} \right) = 0,
\end{equation}
\begin{equation}
\begin{aligned}
 \barrho \left(\imath \kx ( \baru - c ) \tu + \frac{d\baru}{dy} \tv \right)& = -\frac{\imath \kx \tp}{\gamma \Ma^{2}} + \frac{1}{\Re} \Bigg[ \mu\left(\barT\right) \left( \dfrac{d^2}{dy^2} - \kxsq \right) \tu + 
\dfrac{d \mu}{dT}\Bigg|_{T={\barT}}\dfrac{d^2 \baru}{d y^2} \tT \\
 &+ \dfrac{d \mu}{dT}\Bigg|_{T={\barT}}\dfrac{d \barT}{d y} \left( \dfrac{d \tu}{d y} + \imath \kx \tv \right) 
 + \left( \dfrac{d \mu}{dT}\Bigg|_{T={\barT}}\frac{d \tT}{d y} + \dfrac{d^2 \mu}{dT^2}\Bigg|_{T={\barT}}\frac{d \barT}{
 d y} \tT \right) \dfrac{d\baru}{dy} \\
&\:\:\:\:\: + \imath \kx (\mu_b + \mbox{$\frac{1}{3}$}) \mu\left(\barT\right) \left( \imath \kx \tu + \frac{d\tv}{dy}\right) \Bigg],
\end{aligned}
\label{u_mode}
\end{equation}
\begin{equation}
\begin{aligned}
\barrho \imath \kx ( \baru - c ) \tv =& -\frac{1}{\gamma \Ma^{2}} \frac{d\tp}{dy} + \frac{1}{\Re} \Bigg[ \mu\left(\barT\right) 
\left( \dfrac{d^2}{dy^2} - \kxsq \right) \tv + \imath \kx \dfrac{d \mu}{dT}\Bigg|_{T={\barT}}\dfrac{d\baru}{dy} \tT   
+ 2 \dfrac{d \mu}{dT}\Bigg|_{T={\barT}}\dfrac{d \barT}{d y} \dfrac{d\tv}{dy} \\
& + (\mu_b -\frac{2}{3})\dfrac{d \mu}{dT}\Bigg|_{T={\barT}}\dfrac{d \barT}{d y} \left( \imath \kx \tu + \frac{d\tv}{dy} \right) 
+ (\mu_b + \mbox{$\frac{1}{3}$}) \mu\left(\barT\right) \frac{d}{d y} \left( \imath \kx \tu + \frac{d \tv}{dy}\right) \Bigg], 
\label{v_mode}
\end{aligned}
\end{equation}
\begin{equation}
\begin{aligned}
\barrho \left(\imath \kx ( \baru - c ) \tT + \frac{d\barT}{dy} \tv \right) &= - (\gamma - 1) \barrho \barT 
\left( \imath \kx \tu + \frac{d\tv}{dy} \right) 
+ \frac{\gamma}{\Re \Pr} \Bigg[ \kappa\left(\barT\right) \left( \dfrac{d^2}{dy^2} - \kxsq \right)  \tT \\
& + 2 \dfrac{d \kappa}{dT}\Bigg|_{T={\barT}} \dfrac{d \barT}{d y} \dfrac{d \tT}{d y} + \left(\dfrac{d^2 \kappa}{dT^2}\Bigg|_{T={\barT}} \left( \dfrac{d\barT}{dy}\right)^2  + \dfrac{d \kappa}{dT}\Bigg|_{T={\barT}} \dfrac{d^2 \bar{T}}{d y^2} \right) \tT \Bigg] \\
&\:\:+ \dfrac{\gamma (\gamma - 1) \Ma^2}{\Re} \left[ \dfrac{d \mu}{dT}\Bigg|_{T={\barT}} \left( \dfrac{d\bar{u}}{dy}\right)^2 \tT + 2 \mu\left(\barT\right) \dfrac{d\baru}{dy} \left( \dfrac{d \tu}{d y} + \imath \kx \tv \right) \right],
\label{T_mode}
\end{aligned}
\end{equation}
\begin{equation}
\label{p_mode}
    \tp = \barrho \tT + \barT \trho.
\end{equation}
Similarly, the governing equations for the two-dimensional normal modes in the solid are,
\begin{equation}
 \frac{d\tvs}{d\bary} + \imath \kx \tus = 0,
\label{rho_mode_solid}
\end{equation}
\begin{equation}
\begin{aligned}
 -\kxsq c^2 \tus = -\frac{\imath \kx \tp_s}{\gamma \Ma^2} &+ \frac{1}{\lambda}\left( \frac{d^2\tus}{d\bary^2} - \left(1 + \left(\frac{d\barus}{d\bary}\right)^2\right)\kxsq\tus + 2\imath \kx \frac{d\barus}{d\bary}\frac{d\tus}{d\bary} + \imath \kx \frac{d^2\barus}{d\bary^2}\tus \right) \\
	&- \frac{\imath \kx c \mu_r}{\rho_r\Re} \left(\frac{d^2}{d\bary^2} - \kxsq \right)\tus,
\end{aligned}
\label{u_mode_solid}
\end{equation}   
\begin{equation}
\begin{aligned}
 -\kxsq c^2 \tvs = -\frac{1}{\gamma \Ma^2}\frac{d\tp_s}{d\bary} &+ \frac{1}{\lambda}\left( \frac{d^2\tvs}{d\bary^2} - \left(1 + \left(\frac{d\barus}{d\bary}\right)^2\right)\kxsq\tvs + 2\imath \kx \frac{d\barus}{d\bary}\frac{d\tvs}{d\bary} + \imath \kx \frac{d^2\barus}{d\bary^2}\tvs \right) \\
	&- \frac{\imath \kx c \mu_r}{\rho_r\Re} \left(\frac{d^2}{d\bary^2} - \kxsq \right)\tvs,
\end{aligned}
\label{v_mode_solid}
\end{equation} 
\begin{equation}
\begin{aligned}
 -C_r \imath \kx c \tT_s = \frac{\gamma\kappa_r}{\rho_r\Re\Pr}\left( \left(\frac{d^2}{d\bary^2} - \kxsq\right)\tT_s - \frac{d\barT_s}{d\bary}\left(\frac{d^2}{d\bary^2} - \kxsq\right)\tvs - 2 \frac{d^2\barT_s}{d\bary^2} \frac{d\tvs}{d\bary} \right).
\end{aligned}
\label{T_mode_solid}
\end{equation} 

\noindent The boundary conditions for the flow are,
\begin{equation}
 \tu|_{y=1} = \tv|_{y=1} = 0\:,\:\: \tT|_{y=1} = 0, 
\label{eq:bc1_mode}
\end{equation}
and the solid are,
\begin{equation}
\begin{aligned}
& \tus|_{\bary=-L_r} = \tvs|_{\bary=-L_r} = 0\:,\\
& \tT_s|_{\bary=-L_r} = 0 \:(\mbox{for isothermal bottom wall})\:,\:\: \mbox{or}\:,\:\: \frac{d\tT_s}{d\bary}\Big|_{\bary = -L_r} = 0 \:(\mbox{for adiabatic bottom wall}).
\end{aligned}
\label{eq:bc2_mode}
\end{equation}
The interface conditions are,
\begin{equation}
 -\imath \kx c \:\tvs = \tv ,
\label{eq:ic1_mode}
\end{equation}
\begin{equation}
 -\imath \kx c \:\tus  = \tu + \frac{d\baru}{d\bary} \tvs ,
\label{eq:ic2_mode}
\end{equation}
\begin{equation}
 -\frac{\rho_r \Re \tp_s}{\gamma\Ma^2} + \frac{2\rho_r\Re}{\lambda}\left(\frac{d\tvs}{d\bary} + \imath\kx\frac{d\barus}{d\bary}\tvs \right) - {2\imath\kx c \mu_r}\frac{d\tvs}{d\bary} = -\frac{\Re\tp}{\gamma\Ma^2} + {2\barmu}\frac{d\tv}{dy} - \frac{2\barmu}{3}\left(\imath\kx\tu + \frac{d\tv}{dy} \right),
\label{eq:ic3_mode} 
\end{equation}
\begin{equation}
\left(\frac{\rho_r\Re}{\lambda} - {\imath\kx c \mu_r}\right)\left(\frac{d\tus}{d\bary} + \imath\kx\tvs\right) = {\barmu} \left(\frac{d\tu}{dy} + \imath\kx\tv \right) + \frac{d\mu}{dT}\Bigg|_{\barT} \frac{d\baru}{dy} \tT + \frac{d}{dy}\left(\barmu\frac{d\baru}{dy}\right) \tvs ,
\label{eq:ic4_mode}
\end{equation}
\begin{equation}
 \tT_s = \tT + \frac{d\barT}{dy} \tvs ,
\label{eq:ic5_mode}
\end{equation}
\begin{equation}
 \kappa_r \frac{d\tT_s}{d\bary} = \bar{\kappa} \frac{d\tT}{dy} + \frac{d}{dy}\left(\bar{\kappa}\frac{d\barT}{dy}\right) \tvs + \frac{d\kappa}{dT}\Bigg|_{\barT} \frac{d\barT}{dy} \tT.
\label{eq:ic6_mode}
\end{equation}

\section{Asymptotic analysis}

\subsection{Perturbation equations}
\label{secapp:asymp_eqs}

The leading order mass and momentum equations in the solid are,
\begin{equation}
 \frac{d\tvszero}{d\bary} + \imath \kx \tuszero = 0,
\label{eq:rho_mode_solid_highrhor-zero}
\end{equation}
\begin{equation}
 - \kxsq \czerosq \tuszero = -{\imath \kx \tpszero} + \frac{1}{\lambda}\left( \frac{d^2}{d\bary^2} - \kxsq\right)\tuszero,
\label{eq:u_mode_solid_highrhor-zero}
\end{equation}   
\begin{equation}
 -\kxsq \czerosq \tvszero = -\frac{d\tpszero}{d\bary} + \frac{1}{\lambda}\left( \frac{d^2}{d\bary^2} - \kxsq\right)\tvszero,
\label{eq:v_mode_solid_highrhor-zero}
\end{equation} 
which are same as the equations corresponding to a incompressible linear elastic solid. Taking derivative with respect to $\bary$ of Eq. \ref{eq:u_mode_solid_highrhor-zero}, subtracting from $\imath\kx$ times Eq. \ref{eq:v_mode_solid_highrhor-zero}, and substituting $\tuszero = (\imath/\kx)d\tvszero/dy$ from Eq. \ref{eq:rho_mode_solid_highrhor-zero}, we get,
\begin{equation}
\left(D^4 - (2 - \lambda \czerosq) \kxsq D^2 + (1 - \lambda \czerosq)\kx^4 \right) \tvszero  = 0. 
\label{eq:vs_mode_1}
\end{equation}
At the fluid-solid interface, the leading order velocity continuity conditions become,
\begin{equation}
 -\imath \kx \czero \:\tvszero = \tvzero ,
\label{eq:ic1_mode_incomp_highrhor-zero}
\end{equation}
\begin{equation}
 -\imath \kx \czero \:\tuszero  = \tuzero + \frac{d\baru}{dy} \tvszero ,
\label{eq:ic2_mode_incomp_highrhor-zero}
\end{equation}
while the leading order stress continuity conditions are shown in Eqs. \ref{eq:ic3_mode_incomp_highrhor-zero} and \ref{eq:ic4_mode_incomp_highrhor-zero}.

\subsection{Energy equation in the solid}
\label{secapp:ene_solid}

To obtain the leading order fluid eigenfunctions for the elastic modes, the fluid equations (Eqs. \ref{rho_mode}-\ref{T_mode}) are needed to be solved with the velocity interface conditions (Eqs. \ref{eq:ic1_mode_incomp_highrhor-zero} and \ref{eq:ic2_mode_incomp_highrhor-zero}) and a suitable interface condition for temperature. The temperature eigenfunction in the fluid relates to the solid counterpart through the temperature and heat flux continuity conditions (Eqs. \ref{eq:ic5_mode} and \ref{eq:ic6_mode}). Therefore, in order to obtain a temperature interface condition, we need to first obtain the solution for the temperature eigenfunction in the solid.  
The energy equation in the solid, Eq. \ref{T_mode_solid}, can be written as,
\begin{equation}
\begin{aligned}
 - \imath c \beta^2 \tT_s = \epsilon\left( \left(\frac{d^2}{d\bary^2} - \kxsq\right)\tT_s - \frac{d\barT_s}{d\bary}\left(\frac{d^2}{d\bary^2} - \kxsq\right)\tvs - 2 \frac{d^2\barT_s}{d\bary^2} \frac{d\tvs}{d\bary} \right),
\end{aligned}
\label{eq:temp_solid_compact}
\end{equation}  
where, $\rho_r\Re = \epsilon^{-1}$ and $\beta^2 = (\kx C_r\Pr )/(\gamma \kappa_r)$ is a positive quantity. From the equation it is obvious that the leading order solution is trivial (zero) in the bulk of the domain, except across a thermal boundary layer near $\bary = 0$, where the second derivative of temperature is $O(\epsilon)$ (singular perturbation problem). Therefore, we can define a boundary layer co-ordinate, $\eta = \epsilon^{-1/2}\bary$, and re-write Eq. \ref{eq:temp_solid_compact} across the thermal boundary layer as,
\begin{equation}
 -\imath \czero \beta^2 \tTszero = \frac{d^2 \tTszero}{d\eta^2},
\end{equation}
which can be solved with the matching condition, $\lim_{\eta \rightarrow \infty} \tTszero = 0$, to obtain the solution,
\begin{equation}
 \tTszero = A \exp\left(-\left(-\imath\czero\right)^{1/2}\beta \eta \right) = A \exp\left(-\epsilon^{-1/2}\left(-\imath\czero\right)^{1/2}\beta \bary \right)
\end{equation}
The above solution can be substituted into Eqs. \ref{eq:ic5_mode} and \ref{eq:ic6_mode} at $\bary = 0$, to eliminate the constant $A$. The resulting interface condition obtained reads as,
\begin{equation}
-\epsilon^{-1/2} \kappa_r (-\imath\czero)^{1/2} \beta \left( \tTzero + \frac{d\barT}{dy} \tvszero \right) = \bar{\kappa} \frac{d\tTzero}{dy} + \frac{d}{dy}\left(\bar{\kappa}\frac{d\barT}{dy}\right) \tvszero + \frac{d\kappa}{dT}\Bigg|_{\barT} \frac{d\barT}{dy} \tTzero.
\end{equation}
The above Robin condition for temperature can be used as a boundary condition for the temperature eigenfunctions of the fluid. However, an $O(\epsilon^{1/2})$ accurate approximation for the above relation is the simple Dirichlet condition, 
\begin{equation}
 \tTzero + \frac{d\barT}{dy} \tvszero = 0,
\end{equation}
which is what we employ in the asymptotic solutions in this study.

\subsection{Eigenvalue approximation}
\label{secapp:eig_approx}

To obtain the eigenvalue approximation, we require evaluation of the boundary terms in the stability expression, Eq. \ref{eq:tvs_integral}, from the normal and tangential continuity conditions, Eqs. \ref{eq:ic3_mode}-\ref{eq:ic4_mode} (with $\rho_r\Re = \epsilon^{-1}$ substituted). Since the normal stress continuity condition requires the pressure mode, we use the $x$-momentum equation in the solid, Eq. \ref{u_mode_solid} to express $\tp_s$ in terms of $\tvs$. We multiply Eq. \ref{u_mode_solid} by $-\imath \kx$, and use Eq. \ref{rho_mode_solid}, the resulting relation becomes,
\begin{equation}
 \kxsq\tp_s = \kxsq c^2 D\tvs + \frac{1}{\lambda}(D^2 - \kxsq) D\tvs + 2\imath\epsilon \kx \barsigmatw D^2 \tvs - \epsilon\imath \kx c \mu_r (D^2 - \kxsq) D\tvs.
\label{eq:tps_mode}
\end{equation}
At the leading order the above relation can be written as,
\begin{equation}
\lambda \kxsq \tpszero = D^3 \tvszero - (1 - \lambda \czerosq)\kxsq D\tvszero.
\end{equation}
Substituting this into the interface condition, Eq. \ref{eq:ic3_mode_incomp_highrhor-zero}, and simplifying, we get,
\begin{equation}
 D^3 \tvszero = (3 - \lambda \czerosq)\kxsq D\tvszero.
\label{eq:d3tvszero}
\end{equation}
Similarly, using Eq. \ref{rho_mode_solid}, to express $\tuszero$ in terms of $\tvszero$ in the tangential stress continuity condition, Eq. \ref{eq:ic4_mode_incomp_highrhor-zero}, we get,
\begin{equation}
 D^2 \tvszero = - \kxsq \tvszero.
\label{eq:d2tvszero}
\end{equation}
At the leading order, Eq. \ref{eq:tvs_integral} can be written as,
\begin{equation}
\begin{aligned}
\tvsazero D^3 \tvszero \Big|_{-L_r}^0 - &D\tvsazero D^2 \tvszero \Big|_{-L_r}^0 - (2 - \lambda \czerosq)\kxsq \tvsazero D \tvszero \Big|_{-L_r}^0 \\
&+ \int_{-L_r}^0 \left(|D^2 \tvszero|^2 + (2 - \lambda \czerosq)\kxsq |D \tvszero|^2 + (1 - \lambda \czerosq)\kx^4 |\tvszero|^2 \right) d\bary = 0.
\end{aligned}
\end{equation} 
Substituting, $D^3\tvszero$ and $D^2\tvszero$ from Eqs. \ref{eq:d3tvszero} and \ref{eq:d2tvszero}, respectively, for the boundary terms at $y = 0$, and simplifying, we get Eq. \ref{eq:tvs_integral_zero}.  

To derive an expression for the leading order instability, we need to obtain the $O(\epsilon)$ correction to the stability expression Eq. \ref{eq:tvs_integral}. The imaginary part of the $O(\epsilon)$ correction to Eq. \ref{eq:tvs_integral} reads as,
\begin{equation}
\begin{aligned}
\mbox{Imag}\Bigg(\tvsazero D^3 \tvsone \Big|_{-L_r}^0 - D\tvsazero D^2 \tvsone \Big|_{-L_r}^0 - (2 - \lambda \czerosq)\kxsq \tvsazero D \tvsone \Big|_{-L_r}^0 + 2\lambda \czero \cone \kxsq \tvsazero D\tvszero \Big|_{-L_r}^0 \\
+ \tvsaone D^3 \tvszero \Big|_{-L_r}^0 - D\tvsaone D^2 \tvszero \Big|_{-L_r}^0 - (2 - \lambda \czerosq)\kxsq \tvsaone D \tvszero \Big|_{-L_r}^0 \Bigg) \\
 - 2\lambda \cRzero \cIone \kxsq \int_{-L_r}^0 \left(|D \tvszero|^2 + \kxsq |\tvszero|^2 \right) d\bary \\
  - \kx \czero \mu_r \lambda\left( \tvsazero D^3 \tvszero \Big|_{-L_r}^0 - D\tvsazero D^2 \tvszero \Big|_{-L_r}^0 - 2 \kxsq \tvsazero D \tvszero \Big|_{-L_r}^0 + \int_{-L_r}^0  \left(|D^2 \tvszero|^2 + 2\kxsq |D \tvszero|^2 + \kx^4 |\tvszero|^2 \right) d\bary  \right) \\
+ 2 \lambda \kx \barsigmatw \left( \tvsazero D^2 \tvszero \Big|_{-L_r}^0 - \kxsq |\tvsazero|^2 \Big|_{-L_r}^0 - \int_{-L_r}^0 D\tvsazero\left(D^2 - \kxsq\right)\tvszero d\bary  \right)= 0.
\end{aligned}
\label{eq:tvs_integral_one} 
\end{equation}
To evaluate the boundary terms at $y=0$, we, first write the $O(\epsilon)$ equation for the pressure mode from Eq. \ref{eq:tps_mode}, which reads as,
\begin{equation}
 \lambda\kxsq\tpsone = D^3 \tvsone - (1 - \lambda\czerosq)\kxsq D\tvsone + 2\lambda\czero\cone\kxsq D\tvszero + 2\imath\lambda\kx \barsigmatw D^2\tvszero - \imath\kx\lambda\czero\mu_r(D^2 - \kxsq)D\tvszero.
\end{equation}
Substituting into the $O(\epsilon)$ normal stress continuity condition shown in Eq. \ref{eq:ic3_mode_highrhor-first} and simplifying, we get,
\begin{equation}
 D^3\tvsone = (3 - \lambda \czerosq)\kxsq D\tvsone + 4\imath\lambda\kx^3 \barsigmatw \tvszero - 2\lambda\czero\cone\kxsq D\tvszero + \imath\kx\czero\lambda\mu_r(D^2 - 3\kxsq)D\tvszero - \lambda\kxsq \tsigmayyzero ,
\label{eq:d3tvsone}
\end{equation} 
where, $\tsigmayyzero$ is defined in Eq. \ref{eq:mode_str_zero}. Similarly, using the continuity equation, Eq. \ref{rho_mode_solid}, at the first correction order, we can write the $O(\epsilon)$ tangential stress balance condition, Eq. \ref{eq:ic4_mode_highrhor-first}, as,
\begin{equation}
D^2\tvsone = -\kxsq\tvsone - \imath\kx\lambda\left(\tsigmaxyzero + D(\barmu D\baru) \tvszero \right), 
\label{eq:d2tvsone}
\end{equation}
where, $\tsigmaxyzero$ is defined in Eq. \ref{eq:mode_str_zero}. Therefore we substitute Eqs. \ref{eq:d3tvszero}, \ref{eq:d2tvszero}, \ref{eq:d3tvsone} and \ref{eq:d2tvsone} into the boundary terms of Eq. \ref{eq:tvs_integral_one}. Upon algebraic simplification, we obtain Eq. \ref{eq:tvs_integral_zero_simpl}.

\bibliographystyle{unsrt}
\bibliography{ref}

\end{document}